\documentclass[structabstract]{aa} 
\usepackage{hyperref} 
\usepackage{amssymb,amsmath,graphics,multirow,lscape,aas_macros}
\usepackage{graphicx}
\usepackage{natbib}
\bibpunct{(}{)}{;}{a}{}{,}
\usepackage{textcomp}
\usepackage{epsfig}
\usepackage{color}
\usepackage[hang]{subfigure}
\usepackage{txfonts}

\begin{document}

\title{Sulphur abundances in halo giants from the  [\ion{S}{i}] line at 1082 nm and the \ion{S}{i} triplet around 1045 nm\thanks{Based on observations collected at the European Southern Observatory, Chile (ESO program 080.D-0675(A)).}}

\author{H. J\"onsson\inst{1} \and N. Ryde\inst{1} \and P. E. Nissen\inst{2} \and R. Collet\inst{3} \and K. Eriksson\inst{4}  \and M. Asplund\inst{3} \and B. Gustafsson\inst{4}}

\institute{Department of Astronomy and Theoretical Physics, Lund Observatory, Lund University, Box 43, SE-221 00 Lund, Sweden\\ \email{henrikj@astro.lu.se} \and
Department of Physics and Astronomy, Aarhus University, 8000 Aarhus C, Denmark \and
Max Planck Institute for Astrophysics, Karl-Schwarzschild-Strasse 1,  Postfach 1317, 857 41 Garching bei M\"unchen, Germany \and
Department of Physics and Astronomy, Uppsala University, Box 516, SE-751 20 Uppsala, Sweden
}
	    
 \date{Submitted March 9, 2011; accepted April 11, 2011}

\abstract
   {It is still debated whether or not the Galactic chemical evolution of sulphur in the halo followed the constant or flat trend with $[\mathrm{Fe} / \mathrm{H}]$, ascribed to the result of explosive nucleosynthesis in type II SNe. It has been suggested that the disagreement between different investigations of sulphur abundances in halo stars might be due to problems with the diagnostics used, that a new production source of sulphur might be needed in the early Universe, like hypernovae, or that the deposition of supernova ejecta into the interstellar medium is time-delayed.}
   {The aim of this study is to try to clarify this situation by measuring the sulphur abundance in a sample of halo giants using two diagnostics; the \ion{S}{i} triplet around 1045 nm and the [\ion{S}{i}] line at 1082 nm. The latter of the two is not believed to be sensitive to non-LTE effects. We can thereby minimize the uncertainties in the diagnostic used and estimate the usefulness of the triplet in sulphur determination in halo K giants. We will also be able to compare our sulphur abundance differences from the two diagnostics with the expected non-LTE effects in the 1045 nm triplet previously calculated by others.} 
   {High-resolution near-infrared spectra of ten K giants were recorded using the spectrometer CRIRES mounted on VLT. Two standard settings were used; one covering the \ion{S}{i} triplet and one covering the [\ion{S}{i}] line. The sulphur abundances were determined individually with equivalent widths and synthetic spectra for the two diagnostics using tailored 1D model atmospheres and relying on non-LTE corrections from the litterature. Effects of convective inhomogeneities in the stellar atmospheres are investigated.}
   {The sulphur abundances derived from both the 1082 nm [\ion{S}{i}] line and the non-LTE corrected 1045 nm triplet favor a flat trend for the evolution of sulphur. No `high' values of $[\mathrm{S} / \mathrm{Fe}]$ shown in some previous studies are seen in our sample.}
   {We corroborate the flat trend in the $[\mathrm{S} / \mathrm{Fe}]$ vs. $[\mathrm{Fe} / \mathrm{H}]$ plot for halo stars found in other works and cannot find a scatter nor a rise in $[\mathrm{S} / \mathrm{Fe}]$ obtained in some other previous studies. We find the sulphur abundances deduced from the non-LTE corrected triplet  somewhat lower than the abundances from the [\ion{S}{i}] line, possibly indicating too large non-LTE corrections. Considering 3D modeling, however, they might instead be too small. Further we show that the [\ion{S}{i}] line is possible to use as a sulphur diagnostic down to $[\mathrm{Fe} / \mathrm{H}] \sim -2.3$ in giants.}

\maketitle

\section{Introduction} 
Sulphur is an $\alpha$-element (like O, Ne, Mg, Si, Ar, and Ca) and these are believed to be produced mainly in SNe type II by additions of $\alpha$-particles. The SNe type Ia, on the other hand, mainly produce the iron peak elements (Fe, Co, and Ni). Due to the different life times of the two groups of productions sites, the interstellar medium (ISM) in the early Galaxy is believed to contain higher abundance $\alpha$-elements compared to Fe than later on when the SNe type Ia start expelling iron. Thus, stars that form at a certain time serve as markers of the $[\alpha / \mathrm{Fe}]$\footnote{The notation $[\mathrm{A} / \mathrm{B}]=\log \left( \mathrm{N}_{\mathrm{A}} / \mathrm{N}_{\mathrm{B}} \right)_{*}-\log \left( \mathrm{N}_{\mathrm{A}} / \mathrm{N}_{\mathrm{B}} \right)_{\odot}$, where $\mathrm{N}_{\mathrm{A}}$ and $\mathrm{N}_{\mathrm{B}}$ are the number abundances of elements A and B respectively.} ratio in the ISM at that particular stage in the evolution of the Galaxy. This requires that no extra $\alpha$-elements nor Fe has been produced in the star and contaminated the photosphere during the star's life. The theory of nucleosynthesis and stellar structure and evolution suggest this to be the case.
Since the iron abundance, $[ \mathrm{Fe} / \mathrm{H}]$, roughly increases with time, an $[\alpha / \mathrm{Fe}]$ vs. $[\mathrm{Fe} / \mathrm{H}]$ plot is expected to show a plateau for the lowest metallicities and a negative slope from the time when the SNe type Ia start producing large amounts of iron. Regarding sulphur in particular, it is believed to be produced via oxygen burning just like Si and Ca and therefore these three elements are expected to vary in lock-step with each other. The expected flat behavior in the $[\alpha / \mathrm{Fe}]$ vs. $[\mathrm{Fe} / \mathrm{H}]$ plot is however not unambiguously observed for sulphur in all previous works and hence there is no agreement regarding its evolution. This mismatch with the rest of the $\alpha$-elements makes the behavior of sulphur interesting to investigate for its own sake, but the Galactic evolution of sulphur is also important to determine for at least two other reasons:
\begin{itemize}
\item The Galactic evolution of sulphur can be matched against the predictions from modeling of supernovae yields  and thereby testing such models. 
\item The volatile properties of sulphur mean that the gas abundances of sulphur measured in, for example, damped Lyman $\alpha$-systems, believed to be galaxies in making, do not need to be corrected for dust depletion to estimate the sulphur abundance in the system. This abundance together with Zn might be used as a proxy for the $[\alpha / \mathrm{Fe}]$ and thereby the timescale of the star formation history of the system can be inferred \citep{Nissen2004}. In order to make such conclusions possible the Galactic evolution of sulphur needs to be determined first. 
\end{itemize}

Despite its importance, the Galactic evolution of sulphur is not a well studied subject (compared to, e.g., the Galactic evolution of oxygen). This is mainly due to the lack of suitable sulphur diagnostics. During the years several diagnostics have been used with different strengths and weaknesses. Sulphur, being situated directly below oxygen in the periodic table, has a similar set of energy levels and transitions as oxygen. For example the 1082 nm [\ion{S}{i}]  line and the 1045 nm triplet used in this work are analogous to the widely used  [\ion{O}{i}] line at 630 nm and the \ion{O}{i} 845 nm lines.

It turns out that different works, using different sulphur diagnostics, have resulted in several different scenarios for the Galactic evolution of sulphur. From measurements of the 869 nm \ion{S}{i} doublet \citet{Israelian2001} and \citet{Takada-Hidai2002} found a negative slope  in the  $[\mathrm{S} / \mathrm{Fe}]$ vs. $[\mathrm{Fe} / \mathrm{H}]$ plot for all observed iron abundances ($-3\leq [\mathrm{Fe} / \mathrm{H}] \leq +0.5$). To explain this they  proposed either a scenario for sulphur production involving hypernovae \citep[e.g., ][]{Nakamura2001} or time-delayed iron deposition into the ISM as compared to sulphur (see \citet{Ramaty2000} for a description of the mechanism in connection to $[\mathrm{O} / \mathrm{Fe}]$ vs. $[\mathrm{Fe} / \mathrm{H}]$).  A majority of the stars observed by \citet{Israelian2001} and \citet{Takada-Hidai2002} were later re-analyzed using the roughly ten times stronger \ion{S}{i} triplet around 923 nm \citep{Ryde2004,Korn2005}, allowing a more precise abundance determination for metal-poor stars. These new results, contradicting the earlier findings, suggest an evolution analogous to the rest of the $\alpha$-elements. Neither diagnostic is, however, optimal; the 923 nm triplet is stronger than the 869 nm doublet, but on the other hand it is situated in a spectral region heavily plagued with telluric lines and the middle line of the triplet is situated in the wing of the strong Paschen $\zeta$ line.

Further \cite{Takada-Hidai2005} analyzed 22 stars using both the 923 nm triplet and the 869 nm doublet finding a flat trend from both diagnostics, but significant differences in abundances for the stars with $[\mathrm{Fe} / \mathrm{H}] \leq -1.5$. These differences might be due to non-LTE effects in the diagnostics used, and this was investigated by \cite{Takeda2005} by calculating non-LTE corrections for the 869 nm doublet, the 923 nm triplet, and in addition the 1045 nm triplet. They used data for determining the evolution of sulphur from \cite{Israelian2001,Takada-Hidai2002,Ryde2004,Nissen2004,Takada-Hidai2005} and some older works and applied their non-LTE corrections to the 869 nm doublet and the 923 nm triplet (the 1045 nm triplet not being observed at the time). Surprisingly, they found that the discrepancies between the evolution of sulphur measured using the different diagnostics \emph{increased}; the 869 nm doublet suggesting a steady increase of $[\mathrm{S} / \mathrm{Fe}]$ for lower $[\mathrm{Fe} / \mathrm{H}]$ and the 923 nm triplet indicating a plateau for halo stars. To resolve this mismatch and check their non-LTE modeling they proposed observations of the 1045 nm triplet.
 
\citet{Caffau2005} combined spectra from four observation runs, initially aimed at studying other elements into the largest sample of Galactic sulphur abundance measurements with a total of 74 dwarfs. They used  the NTT and VLT telescopes and since the wavelength coverage of the observation runs varied, they did not use the same diagnostics in all their sulphur abundance determinations. For each star they used as many diagnostics as possible of a very weak multiplet around 675 nm, the doublet around 869 nm, and the triplet around 923 nm. They found an $[\mathrm{S} / \mathrm{Fe}]$ vs. $[\mathrm{Fe} / \mathrm{H}]$ plot with a large scatter in $[\mathrm{S} / \mathrm{Fe}]$ for $-2.4 \leq [\mathrm{Fe} / \mathrm{H}] \leq -1$ resembling a combination of the two types of evolution previously described.

The development of better infrared spectrometers has made more sulphur lines measurable and the problem with finding a suitable diagnostic for determining sulphur abundance is nowadays less severe. One example is the [\ion{S}{i}] line at 1082 nm used in this work. This line is not believed to be affected by non-LTE effects, but unfortunately the [\ion{S}{i}] line is undetectable in halo dwarfs for low $[\mathrm{Fe} / \mathrm{H}]$. It is however detectable in giants down to $[\mathrm{Fe} / \mathrm{H}]  \sim -2.5$. Another example is the triplet around 1045 nm also used in this work. This triplet is not as strong as the triplet at 923 nm, but it is situated in a spectral region almost unaffected by telluric lines. \citet{Caffau2007b} investigated the possibility of using the 1045 nm triplet for sulphur determination in disk dwarfs and it was recently used by them for determining sulphur abundances in four halo dwarfs corroborating the scatter in $[\mathrm{S} / \mathrm{Fe}]$ for $[\mathrm{Fe} / \mathrm{H}] \sim -1$ found in \citet{Caffau2005} but at a lower level \citep{Caffau2010}. 

\cite{Takeda2010} also recently used the 1045 nm triplet to determine $[\mathrm{S} / \mathrm{Fe}]$ in 33 halo/disk stars proposing yet a new scenario for the evolution of sulphur; a zig-zag trend with a local plateau around $[\mathrm{S} / \mathrm{Fe}] \sim 0.3$ for $-2.5 \leq [\mathrm{Fe} / \mathrm{H}] \leq -1.5$ preceded by a rise in $[\mathrm{S} / \mathrm{Fe}]$ for lower $[\mathrm{Fe} / \mathrm{H}]$. This is however not confirmed by \citet{Spite2011} using the 923 nm triplet to determine sulphur abundance in 33 giants and turnoff stars, with iron abundances $[\mathrm{Fe} / \mathrm{H}] \leq -2.5$, without finding any stars with `high' $[\mathrm{S} / \mathrm{Fe}]$.

Obviously there is no consensus on the Galactic chemical evolution of sulphur. Is there a scatter or some kind of rise for halo stars or not? In this work we will present and compare sulphur abundance measurements for 10 halo K giants using both the [\ion{S}{i}] line at 1082 nm and the 1045 nm triplet. Since the [\ion{S}{i}] line is not affected by non-LTE effects we will also be able to estimate the non-LTE effects of the 1045 nm triplet by comparing the derived sulphur abundances from the two diagnostics. Thereby we will be able to check the validity of the non-LTE corrections calculated by \cite{Takeda2005} for halo K giants. Since non-LTE corrections seem to be significant for most sulphur diagnostics this empirical test of the corrections is important.

\section{Observations}

We have observed ten K giants in the Galactic halo using the spectrometer CRIRES \citep{Kaufl2004,Moorwood2005,Kaufl2006} mounted on VLT. Giants were chosen because the 1082 nm [\ion{S}{i}] line is too weak to be observed in metal-poor dwarfs (see Sect. \ref{blends}). CRIRES is a high-resolution echelle spectrometer designed for near-infrared observations and it uses nodding and jittering to eliminate the sky background and adaptive optics to enhance the S/N. Basic data for the observed stars are shown in Table \ref{tab:starinfo} and a summary of the observations is shown in Table \ref{tab:criresobs}. The difference in S/N in the two settings is in some cases due to different integration times and in changes due to clouds in the star's visibility between the two observations.

\begin{table*}[htp]
\caption{Basic data for the observed stars.}
\begin{tabular}{l l c c c c c}
\hline
\hline
\multicolumn{2}{c}{Star identifier} & RA (J2000)\tablefootmark{1} & Dec (J2000)\tablefootmark{1} & V\tablefootmark{1} & B-V\tablefootmark{1} & Parallax\tablefootmark{1}\\
 & & (h m s) & (d am as)& & & (mas)\\
\hline
HD13979   	& HIP10497 	& 02 15 20.8536 	& -25 54 54.861  	& 9.17 	& $0.647 \pm 0.024$ 	& $1.93 \pm 1.25$  \\
HD21581   	& HIP16214 	& 03 28 54.4853 	& -00 25 03.117  	&  8.70 	& $0.786 \pm 0.018$	& $4.27 \pm 1.20$  \\ 
HD23798   	& HIP17639 	& 03 46 45.7217 	& -30 51 13.329  	&  8.28 	& $1.033 \pm 0.015$	& $0.71 \pm 0.95$  \\
HD26297   	& HIP19378 	& 04 09 03.4175 	& -15 53 27.068  	&  7.46 	& $1.088 \pm 0.011$ 	&  $1.28 \pm 1.01$  \\
HD29574   	& HIP21648 	& 04 38 55.7328 	& -13 20 48.138  	&  8.33 	& $1.304 \pm 0.019$	& $0.66 \pm 1.03$  \\
HD36702   	& HIP25916 	& 05 31 52.2305 	& -38 33 24.046  	&  8.33 	& $1.148 \pm 0.018$ 	&  $0.68 \pm 0.78$  \\
HD44007   	& HIP29992 	& 06 18 48.5269 	& -14 50 43.424  	& 8.05 	& $0.829 \pm 0.012$ 	& $5.17 \pm 1.02$  \\
HD83212   	& HIP47139 	& 09 36 19.9533 	& -20 53 14.759  	&  8.33 	& $1.015 \pm 0.015$	& $1.96 \pm 0.98$  \\
HD85773   	& HIP48516 	& 09 53 39.2415 	& -22 50 08.425  	&  9.42 	& $1.078 \pm 0.046$ 	& $4.07 \pm 1.30$  \\
HD103545 	& HIP58139 	& 11 55 27.1618 	& +09 07 45.028 	&  9.42 	& $0.844 \pm 0.043$ 	& $0.60 \pm 1.23$  \\
\hline
\end{tabular}
\label{tab:starinfo}
\tablebib{(1)~\citet{Hip}}
\end{table*}

\begin{table}[htp]
\caption{Summary of the observations with CRIRES/VLT.}
\begin{tabular}{l c c c c c}
\hline
\hline
Star & Date & \multicolumn{2}{c}{Integration time\tablefootmark{a}} & \multicolumn{2}{c}{$\mathrm{S}/\mathrm{N}$\tablefootmark{b}} \\
& & Triplet &  [\ion{S}{i}] & Triplet &  [\ion{S}{i}] \\
 & & (s) & (s) \\
\hline
HD13979 		&  30 Oct. 2007 	&  1080 	& 1440	& 360 	& 570\\
HD21581 		&  16 Jan. 2008 	&  960 	& 960	& 330 	& 310\\ 
HD23798 		&  31 Oct. 2007 	&  840 	& 840	& 500 	& 520\\
HD26297 		& 16 Jan. 2008 	&  240 	& 240 	& 200	& 330\\
HD29574 		& 16 Jan. 2008 	&  360 	& 360	& 300 	& 310\\
HD36702 		& 16 Jan. 2008 	&  720 	& 720 	& 320 	& 360\\
HD44007 		& 15 Feb. 2008 	&  480 	& 480 	& 370 	& 350\\
HD83212 		& 16 Jan. 2008 	&  480 	& 480	& 310 	& 240\\
HD85773 		& 16 Jan. 2008 	&  1080 	& 1080 	& 480  	& 440\\
HD103545 	& 16 Jan. 2008 	&  1440 	& 1080	& 110 	& 160\\
\hline
\end{tabular}
\label{tab:criresobs}
\tablefoot{\\
\tablefootmark{a}{The total integration times are given by NDIT $\times$ DIT $\times$ NEXP $\times$ NABCYCLES $\times$ 2, see the CRIRES User Manual at \href{http://www.eso.org/sci/facilities/paranal/instruments/crires/doc/}{http://www.eso.org/sci/facilities/paranal/instruments/crires/doc/}\\}
\tablefootmark{b}{Approximate signal-to-noise ratios per pixel of the observed spectra.}
}
\end{table}

The observations were made with a slit width of 0.25" resulting in a spectral resolution of $R= \lambda / \Delta \lambda \sim 80000$ and  2.5 pixels per resolution element. They were carried out in service mode during October 2007 - February 2008. We used two standard settings: one covering the \ion{S}{i} triplet around 1045 nm and one covering the [\ion{S}{i}] line at 1082 nm. The spectral range of the two settings used are roughly 12 nm with the sulphur lines as centered as possible.

A couple of fast rotating B stars were also observed used for checking that no telluric lines were affecting the lines analyzed.

\subsection{Blending lines and stellar sample}\label{blends}
\begin{figure*}[ht]
\centering
\includegraphics{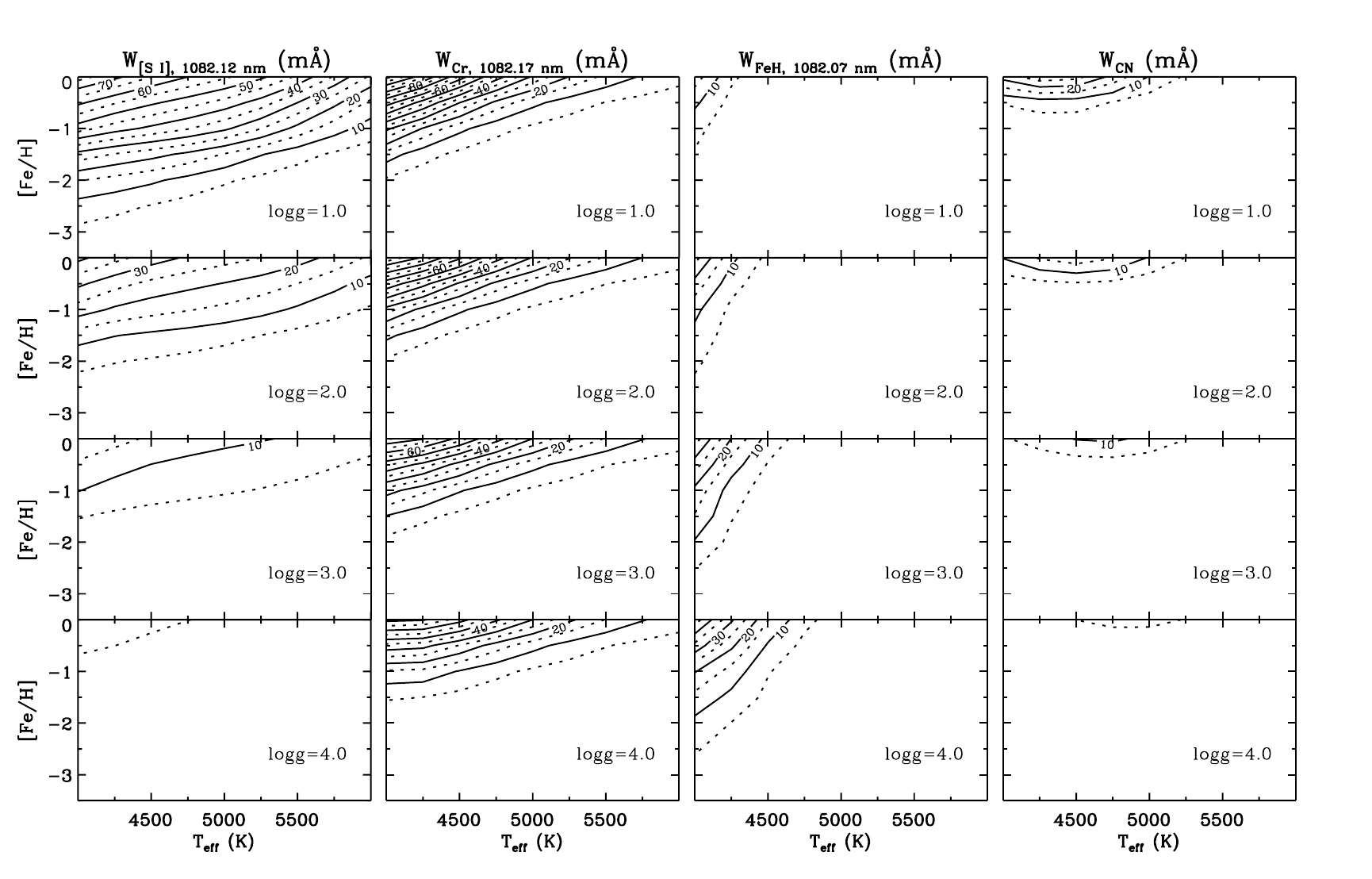}
\caption{Equivalent widths of the 1082 nm [\ion{S}{i}] line and potential blends for different stellar parameters. The leftmost panel shows the sum of the three CN lines listed in Table \ref{tab:blends}. The filled contours mark steps of 10 m\AA  \;and the dotted contours mark the steps of 5 m\AA  \;in between.}
\label{fig:SI_eqw}
\end{figure*}

\begin{figure*}[ht]
\centering
\includegraphics{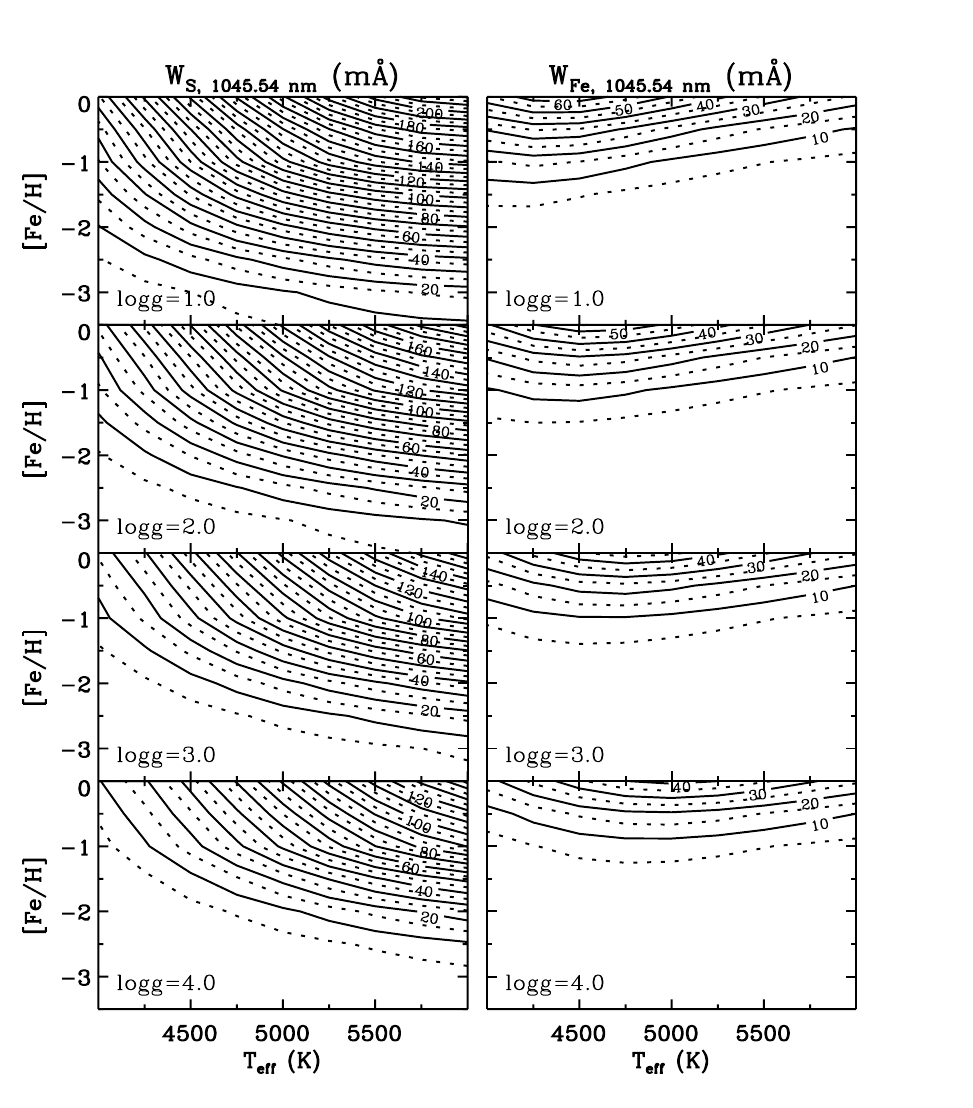}
\caption{Equivalent widths of the 1045.5 nm \ion{S}{i} triplet line and its potential blends for different stellar parameters. The filled contours mark steps of 10 m\AA  \;and the dotted contours mark the steps of 5 m\AA  \;in between.}
\label{fig:Sa_eqw}
\end{figure*}

\begin{figure*}[ht]
\centering
\includegraphics{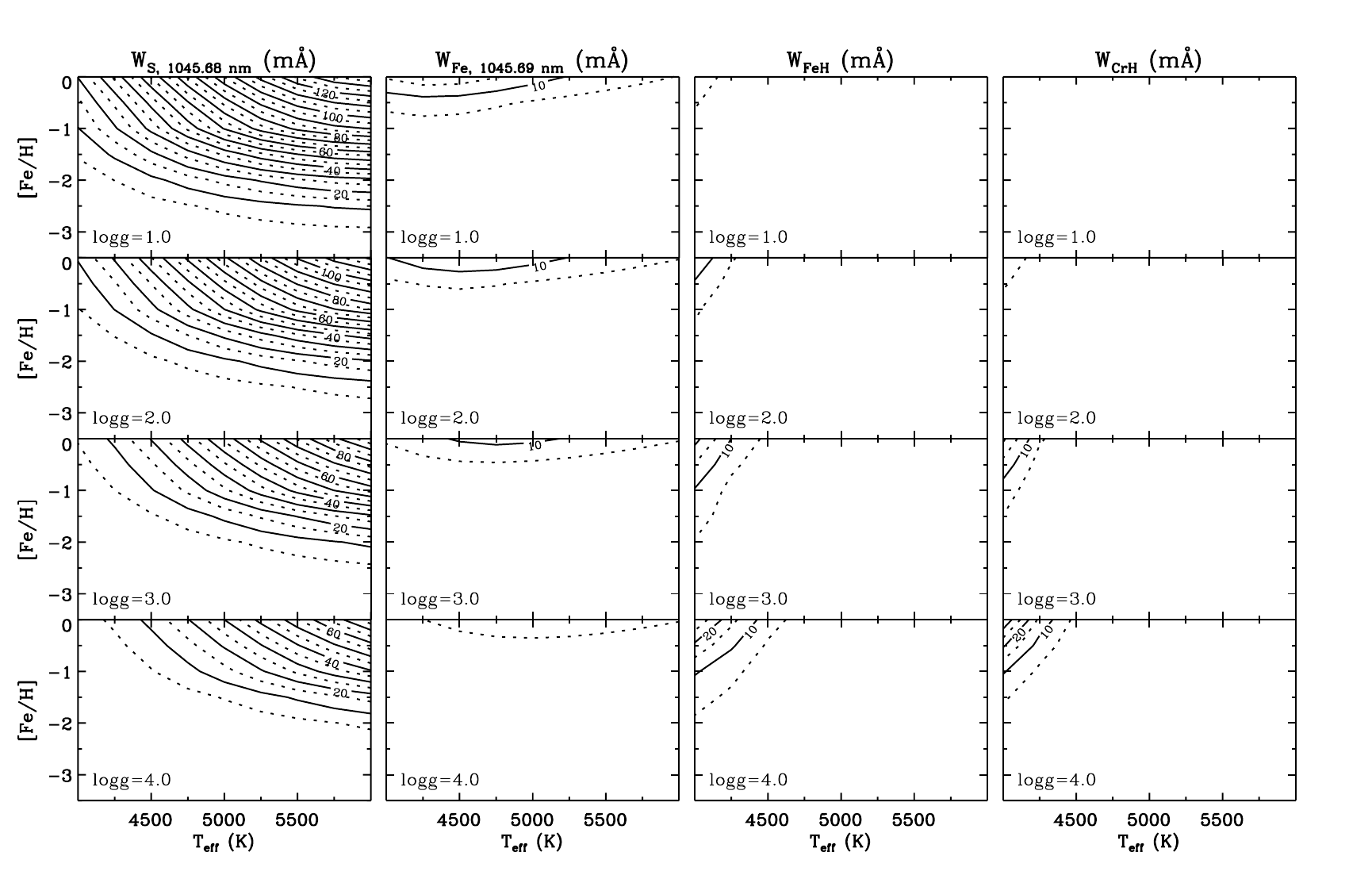}
\caption{Equivalent widths of the 1045.7 nm \ion{S}{i} triplet line and its potential blends for different stellar parameters. The two leftmost panels show the sum of the three FeH lines and the two CrH lines listed in Table \ref{tab:blends}. The filled contours mark steps of 10 m\AA  \;and the dotted contours mark the steps of 5 m\AA  \;in between.}
\label{fig:Sb_eqw}
\end{figure*}

Using only a few lines, or even just one, for an abundance determination such as ours calls for an extra careful examination of blends, and in case blends are found: to try to choose the stellar sample in such a way that the effects of the blending are minimized. The [\ion{O}{i}] line at 630 nm, analogous to our [\ion{S}{i}] line, is for example blended with a \ion{Ni}{i} line, but the blending is not relevant for metal-poor stars \citep{Allende2001}. To determine for which of our sulphur lines there might be significant blends we have calculated synthetic equivalent widths for all lines in the relevant wavelength region in a grid of 288 model atmospheres. For metals we have used a line list of a relevant wavelength section from the VALD I database \citep{Kupka1999} updated according to \cite{Gustafsson2008}. The lists of molecules we were provided by Bengt Edvardsson (private communication), in turn mostly compiled by Bertrand Plez, and they include CH \citep{Jorgensen1996}, OH \citep{Goldman1998}, CrH \citep{Burrows2002}, SiH (electronic Kurucz\footnote{\href{http://kurucz.harvard.edu/}{http://kurucz.harvard.edu/}}), FeH \citep{Dulick2003}, H$_2$O \citep{Barber2006}, C$_2$ \citep{Querci1971,Querci1974}, CaH, CN, TiO, VO, and, ZrO (electronic and unpublished Plez\footnote{\href{http://www.graal.univ-montp2.fr/hosted/plez/}{http://www.graal.univ-montp2.fr/hosted/plez/}}).  The list of lines with equivalent widths greater than zero within $\pm 0.05$ nm ($\Delta v \sim 30$ km/s) from the used sulphur lines is presented in Table \ref{tab:blends}. The equivalent widths of the 1082 nm [\ion{S}{i}], the 1045.5 nm and 1045.7 nm triplet lines, and blends with strengths in parity with the relevant sulphur line  are shown in Figs. \ref{fig:SI_eqw} - \ref{fig:Sb_eqw}. The 1045.9 nm triplet line does not have any known signiÞcant blends.

\begin{table}[htp]
\caption{Potential blending lines turning up in our grid of models.}
\begin{tabular}{l c c c}
\hline
\hline
Element  	&  Wavelength 	& $\chi_{\mathrm{exc}}$ 	& $\log (gf)$\\
 		&  (nm) (air) 	& (eV )				&\\
\hline
\multicolumn{4}{l}{Blends for the 1082.1176 nm [\ion{S}{i}] line}\\
\hline

\ion{Cr}{i}			& 1082.1658 	& 3.013 	& -1.678\\
\ion{Fe}{i} 			& 1082.0882	& 9.760	& -1.470\\
FeH				& 1082.0683 	& 0.619	& -0.331\\
$^{12}$C$^{12}$C	& 1082.0866 	& 0.690	& -0.888\\
$^{12}$C$^{12}$C	& 1082.1215 	& 0.610	& -1.316\\
$^{12}$C$^{14}$N	& 1082.0779 	& 1.908	& -0.845\\
$^{12}$C$^{14}$N	& 1082.0952 	& 1.552	& -1.379\\
$^{12}$C$^{14}$N	& 1082.1382 	& 2.143	& -1.319\\
\hline
\multicolumn{4}{l}{Blends for the 1045.5449 nm \ion{S}{i} line}\\
\hline
\ion{P}{i}			& 1045.5870	& 8.080	& 0.000\\
\ion{Fe}{i}			& 1045.5404 	& 5.390	& -0.905\\
$^{12}$C$^{12}$C	& 1045.5504	& 1.370	& -1.233\\
$^{12}$C$^{12}$C	& 1045.5614	& 0.340	& -1.699\\
CrH				& 1045.5303	& 0.600	& -0.698\\
TiO				& 1045.5562	& 1.240	& -0.006\\
TiO				& 1045.5583	& 1.170	& -0.344\\
\hline
\multicolumn{4}{l}{Blends for the 1045.6757 nm \ion{S}{i} line}\\
\hline
\ion{P}{i}			& 1045.6836	& 8.080	& -1.000\\
\ion{Ca}{i}			& 1045.7098	& 4.740	& -1.741\\
\ion{Cr}{i} 			& 1045.6267	& 4.420	& -1.855\\
\ion{Cr}{i} 			& 1045.6467	& 6.800	& -2.608\\
\ion{Mn}{i}	 		& 1045.6346	& 5.820	& -1.210\\
\ion{Fe}{i}			& 1045.6455	& 6.200	& -1.479\\
\ion{Fe}{i}			& 1045.6667	& 4.190	& -4.669\\
\ion{Fe}{i}			& 1045.6939	& 5.540	& -1.561\\
$^{12}$C$^{12}$C	& 1045.6636	& 0.250	& -1.854\\
$^{12}$C$^{12}$C	& 1045.6731	& 1.370	& -1.242\\
$^{12}$C$^{13}$C	& 1045.7016	& 0.270	& -1.345\\
$^{12}$C$^{12}$C	& 1045.7133	& 0.620	& -1.662\\
$^{12}$C$^{14}$N	& 1045.6394	& 1.160	& -2.931\\
$^{13}$C$^{14}$N	& 1045.6915	& 1.160	& -1.947\\
$^{13}$C$^{14}$N	& 1045.7044	& 1.530	& -0.864\\
$^{12}$C$^{14}$N	& 1045.7253	& 1.760	& -1.346\\
CrH				& 1045.6629	& 0.078	& -0.728\\
CrH				& 1045.6983	& 0.440	& -0.740\\
FeH				& 1045.6292	& 0.100	& -2.539\\
FeH				& 1045.6379	& 0.710	& -2.576\\
FeH				& 1045.7003	& 0.470	& -0.869\\
TiO				& 1045.6596	& 1.390	& 0.352\\
TiO				& 1045.6653	& 1.270	& 0.042\\
TiO				& 1045.7061	& 1.210	& -0.393\\
\hline
\multicolumn{4}{l}{Blends for the 1045.9406 nm \ion{S}{i} line}\\
\hline
\ion{V}{i}			&1045.9169	& 3.070	& -1.719\\
$^{12}$C$^{12}$C	& 1045.8992	& 0.620	& -1.672\\
$^{12}$C$^{12}$C	& 1045.9890	& 0.410	& -1.253\\
$^{12}$C$^{14}$N	& 1045.9240	& 1.330	& -2.576\\
FeH				& 1045.9672	& 0.220	& -3.360\\
TiO				& 1045.8933	& 1.510	& 0.051\\
TiO				& 1045.9259	& 1.210	& -0.074\\
\hline
\end{tabular}
\label{tab:blends}
\end{table}

The only blend for the [\ion{S}{i}] line present in halo giants is that due to \ion{Cr}{i}, but this line is separated enough from the sulphur line to be resolvable in our high resolution spectra, see for example the \ion{Cr}{i} line in Fig. \ref{fig:allspectra}. For the 1045.5 nm and 1045.7 nm \ion{S}{i} triplet lines the 1045.54 nm and 1045.69 nm \ion{Fe}{i} lines shown in Fig. \ref{fig:Sa_eqw} and Fig. \ref{fig:Sb_eqw} are close enough to the relevant sulphur line that they would be indistinguishable.

Our sample stars all have stellar parameters available in the literature ($T_{\mathrm{eff}}$, $\log g$, $\left[\mathrm{Fe}/\mathrm{H}\right]$, and  $\xi_{\mathrm{micro}}$). They are chosen to be cool giants in order to maximize the strength of the [\ion{S}{i}] line. Since we are interested in halo stars with $[\mathrm{Fe} / \mathrm{H}] \le -1$ the unresolvable \ion{Fe}{i} blends expected to most seriously affect the 1045 nm sulphur lines are small compared to the relevant sulphur lines and can safely be ignored. Our lines can thus be considered blend free and the measured equivalent widths are that of the sulphur lines only. 

Note that, as can be seen in Figs. \ref{fig:SI_eqw} - \ref{fig:Sb_eqw} one should use these sulphur lines with caution for certain stellar parameters; e.g. for a solar-metallicity dwarf with a temperature of 4500 K the strength of the blending \ion{Fe}{i} line would be roughly equal to that of the 1045.5 nm \ion{S}{i} triplet line. Also, the neighboring \ion{Cr}{i} and [\ion{S}{i}] lines might become unresolvable and therefore blend when observing at lower resolution.

\subsection{Reduction of the spectra}
The data were originally reduced using the CRIRES pipeline version 1.5.0\footnote{Described in the CRIRES Pipeline User Manual, Issue 1.0 at \href{ftp://ftp.eso.org/pub/dfs/pipelines/crire/crire-pipeline-manual-1.0.pdf}{ftp://ftp.eso.org/pub/dfs/pipelines/crire/crire-pipeline-manual-1.0.pdf}}. The pipeline makes use of and automatically handles dark frames, flat fields from a halogen lamp and observations of a ThAr lamp for wavelength calibration.
To subtract the sky background the telescope was nodded between two positions along the slit during the observations. The two frames are subtracted from each other producing two stellar spectra in vertically different places on the detector arrays. The pixels are subsequently added column wise in the slit direction and extracted for each spectrum exposure by the pipeline and then added to produce the final spectrum of the star.

\begin{figure*}[ht]
\subfigcapmargin=4pt
\subfigure[The flat field from 17 jan 2008 of detector two (of four) for the 1082 nm-setting. The optical ghost is clearly visible as a broad band.]{
\includegraphics[width=55mm]{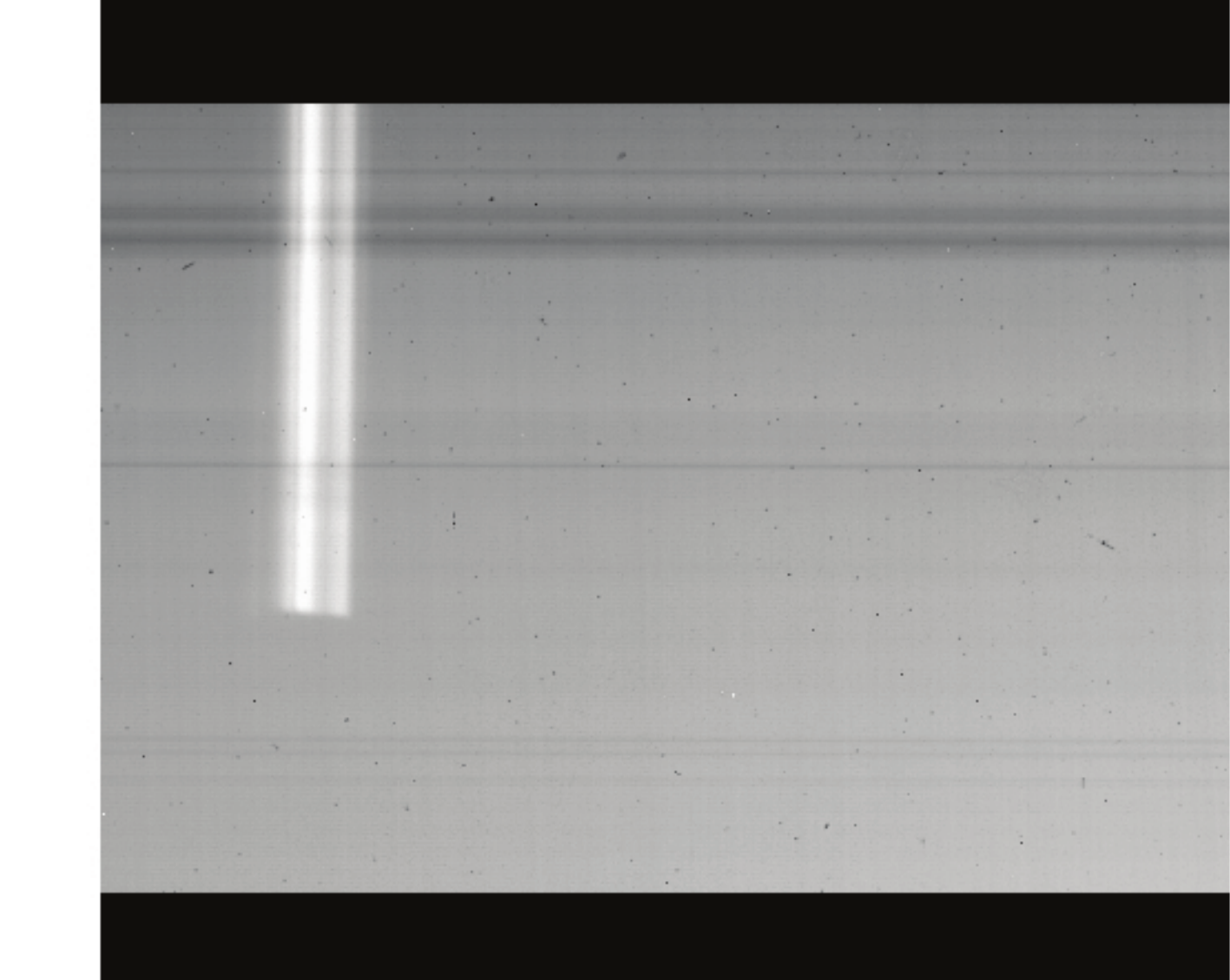}
\label{fig:ccd1082}
}
\subfigure[The flat field from 17 jan 2008 of detector two (of four) for the 1045 nm-setting. The optical ghost is clearly visible as a broad band in another location on the detector.]{
\includegraphics[width=55mm]{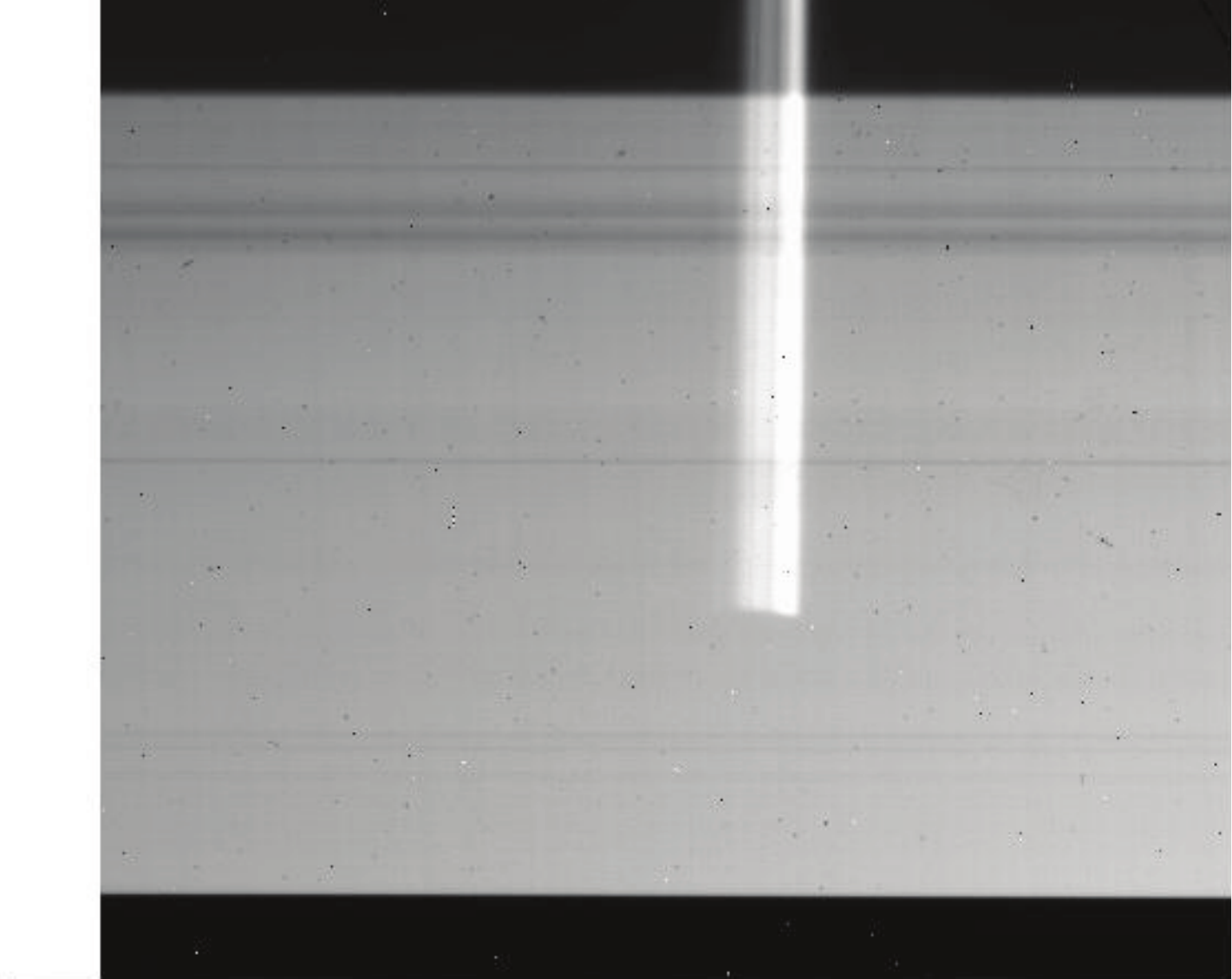}
\label{fig:ccd1045}
}
\subfigure[The nodding of the spectrometer results in two observations of the same star. The raw frames are here added to show both in one picture. The upper observation is affected by the flat field ghost but not the lower. The ghost can still be seen as the dots in the upper left-hand corner. The vertical line is in this case a telluric emission line.]{
\includegraphics[width=55mm]{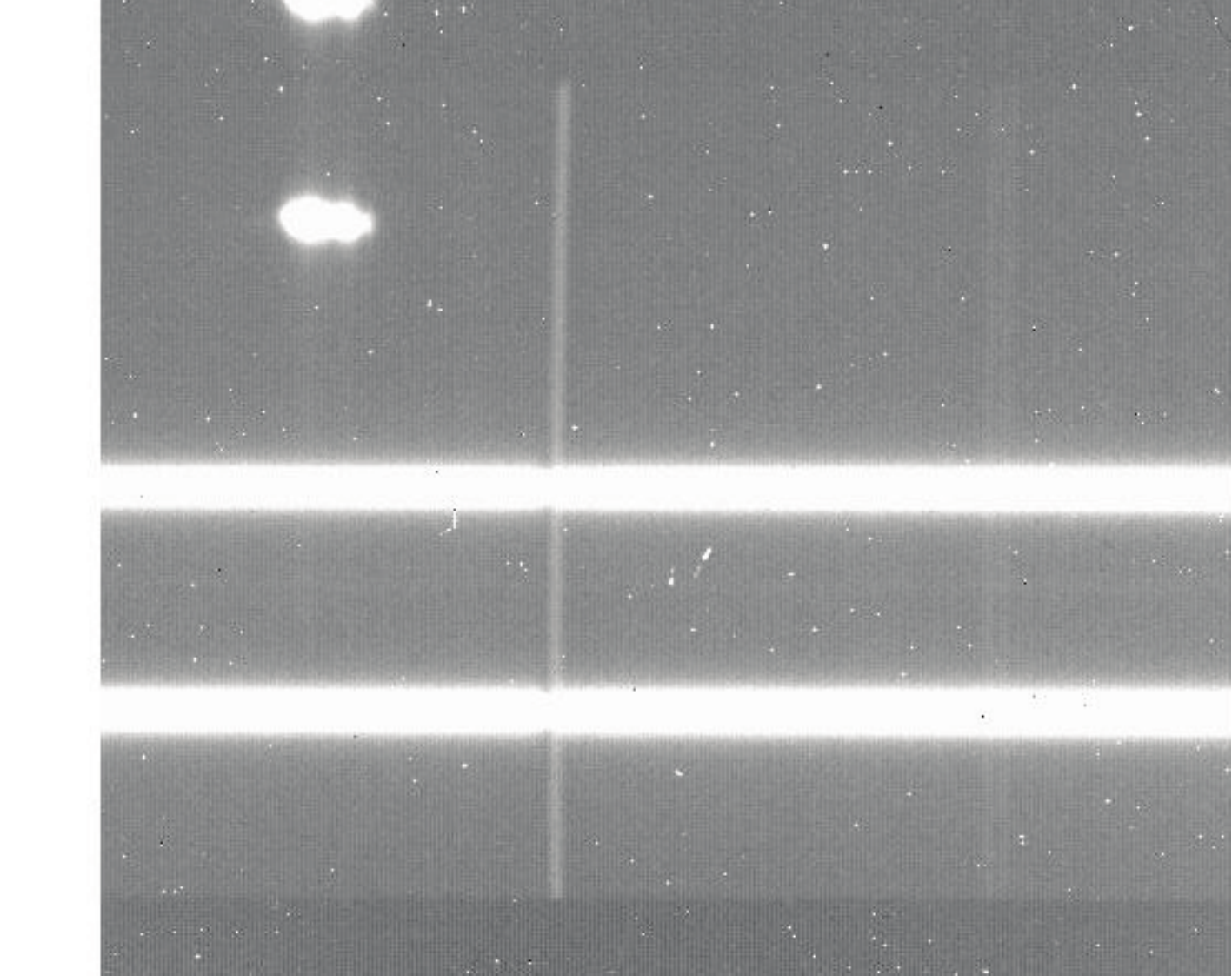}
\label{fig:ccdnodding}
}
\\
\subfigure[The data for HD36702 reduced with pipeline showing the ghost-feature at 1082 nm.]{
\includegraphics[width=60mm]{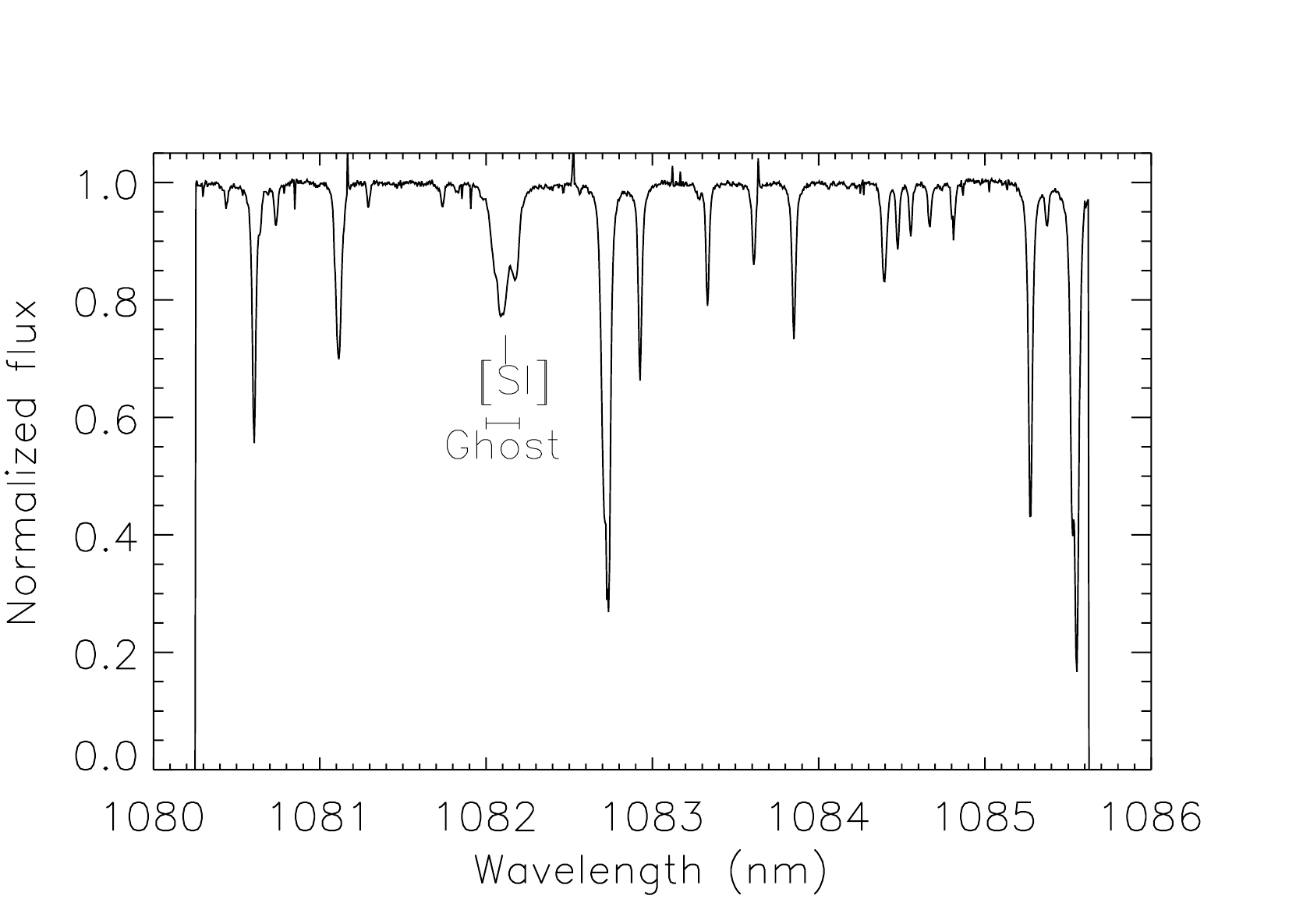}
\label{fig:pipeline}
}
\subfigure[The data for HD36702 re-reduced using a flat field from 1047 nm showing the ghost-feature at 1083 nm.]{
\includegraphics[width=60mm]{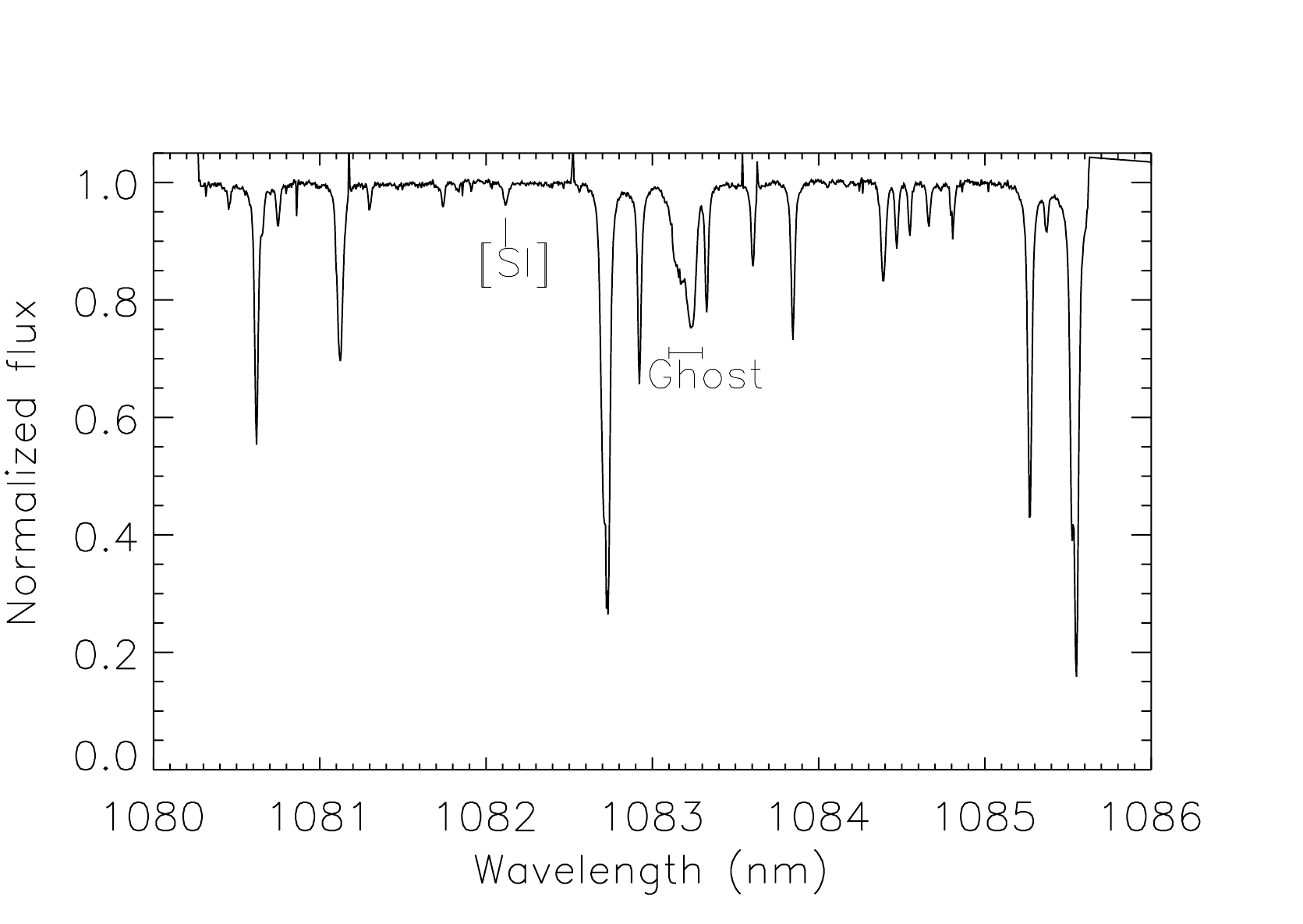} 
\label{fig:rereduced}
}
\caption{The optical ghost of CRIRES and its effects on reduced spectra.}
\label{fig:ghost}
\end{figure*}

Unfortunately there is an optical ghost\footnote{Described in the CRIRES User Manual at \href{http://www.eso.org/sci/facilities/paranal/instruments/crires/doc/}{http://www.eso.org/sci/facilities/paranal/instruments/crires/doc/}} in certain settings of the spectrometer, clearly visible as broad vertical bands in our flat fields (see Fig. \ref{fig:ccd1082} and \ref{fig:ccd1045}), but also as smaller dots in our science frames (see Fig. \ref{fig:ccdnodding}, where a smaller version of the ghost shown in Fig. \ref{fig:ccd1082} can be seen). The optical ghost is present in some higher order settings for the echelle grating and is a result of retro-reflection from the detector to the grating which is redirected onto the detector in a different order.

The ghosts in the flat fields lie in such a location on the detector that it will affect one of the two nod positions (compare the flat fields in Fig. \ref{fig:ccd1082} and \ref{fig:ccd1045} with the stellar spectra in Fig. \ref{fig:ccdnodding}). This results in a spectral artifact in the two spectral settings, most severely affecting our analysis of the [\ion{S}{i}] line; by chance the spectral artifact lies right on top of the [\ion{S}{i}] line in the pipeline reduced spectra making it impossible to measure (see Fig. \ref{fig:pipeline}). Therefore the data covering the [\ion{S}{i}] line had to be re-reduced. We did this using the CRIRES pipeline version 1.11.0\footnote{Described in the CRIRES Pipeline User Manual, Issue 1.4 at \href{ftp://ftp.eso.org/pub/dfs/pipelines/crire/crire-pipeline-manual-1.4.pdf}{ftp://ftp.eso.org/pub/dfs/pipelines/crire/crire-pipeline-manual-1.4.pdf}} using the flat field from the triplet setting instead, which was taken close in time. The spectral feature from the ghost is then situated in another part of the spectra not affecting our analysis (see Fig. \ref{fig:rereduced}). We found that this way of rescuing the data was better than removing the ghost feature by dividing by a telluric star or by reducing the spectra without flat fields, thereby not introducing the ghost, and then dividing the spectra with a B star to remove the "pixel-to-pixel" variations in the detector. The latter method was used successfully by \citet{Nissen2007} when analysing science-verification spectra from CRIRES without flat fields. Our measured S/N values (per pixel of the spectra) of the reduced spectra reveal that our method indeed produces satisfactory signal-to-noise as shown in Table \ref{tab:criresobs}.

The most relevant parts of the reduced spectra can be seen in Fig. \ref{fig:allspectra}. We normalized the spectra and fitted the continua by using the \texttt{IRAF}\footnote{IRAF is distributed by the National Optical Astronomy Observatories, which are operated by the Association of Universities for Research in Astronomy, Inc., under cooperative agreement with the National Science Foundation.} task \texttt{continuum} \citep{Tody1993}.

\subsection{Equivalent widths} \label{eqwi}
The equivalent widths of the 1045 nm \ion{S}{i} triplet and the 1082 nm [\ion{S}{i}] line were measured using the \texttt{splot} task in \texttt{IRAF} by fitting a Gaussian profile to the line, or, in some cases, by straight numerical integration. For two stars (HD26297 and HD29574) the equivalent widths for the forbidden line were measured by de-blending due to the neighboring telluric line. The measured values are listed in Table \ref{tab:eqw}. Since our spectra have high S/N the uncertainties in the measured equivalent widths are mainly due to uncertainties in the continuum setting. The quoted uncertainties have been estimated by fitting the continuum to achieve a maximum and minimum possible value for the equivalent width. The uncertainties in the measured equivalent widths will in the end produce uncertainties in the $\left[\mathrm{S}/\mathrm{Fe}\right]$ less than the uncertainties introduced due to uncertainties in the stellar parameters (see Sect. \ref{stellarparams}). 

\begin{table}[htp]
\caption{Measured equivalent widths for the observed sulphur lines.}
\begin{tabular}{l c c c c}
\hline
\hline
Star 	& 1045.5 nm 	& 1045.7 nm 	& 1045.9 nm 	& 1082.1 nm\\
	& (m\AA) 		& (m\AA) 		& (m\AA) 		& (m\AA)\\
\hline
HD13979   	& 18 $\pm$ 1		& 4 $\pm$ 1       	& 12 $\pm$ 1     	& $\leq 1$ \\
HD21581   	& 48 $\pm$ 1     	& 16 $\pm$ 0.5  	& 32 $\pm$ 0.5 	& 6 $\pm$ 1\\    
HD23798   	& 28 $\pm$ 0.5  	& ...                      	& 18 $\pm$ 0.5 	& 10 $\pm$ 0.5\\
HD26297   	& 39 $\pm$ 1     	& 12 $\pm$ 1     	& 26 $\pm$ 1     	& 16 $\pm$ 1\\
HD29574   	& 41 $\pm$ 1     	& 12$\pm$ 1      	& 28 $\pm$ 1     	& 19 $\pm$ 0.5\\
HD36702   	& 27 $\pm$ 0.5  	& 6 $\pm$ 0.5    	& 19 $\pm$ 0.5 	& 13 $\pm$ 1\\
HD44007   	& 46 $\pm$ 0.5  	& 15 $\pm$ 0.5  	& 32 $\pm$ 0.5 	& 4 $\pm$ 1\\
HD83212   	& 65 $\pm$ 1     	& 25 $\pm$ 0.5  	& 47 $\pm$ 0.5 	& 17 $\pm$ 1\\
HD85773   	& 16 $\pm$ 0.5  	& ...                      	& 12 $\pm$ 0.5     	& 8 $\pm$ 0.5\\
HD103545 	& 26 $\pm$ 2     	& ...                     		& 14 $\pm$ 2      	& $\leq 5$\\
\hline
\end{tabular}
\label{tab:eqw}
\end{table}

\section{Analysis}
\begin{figure*}[ht]
\centering
\includegraphics[width=140mm]{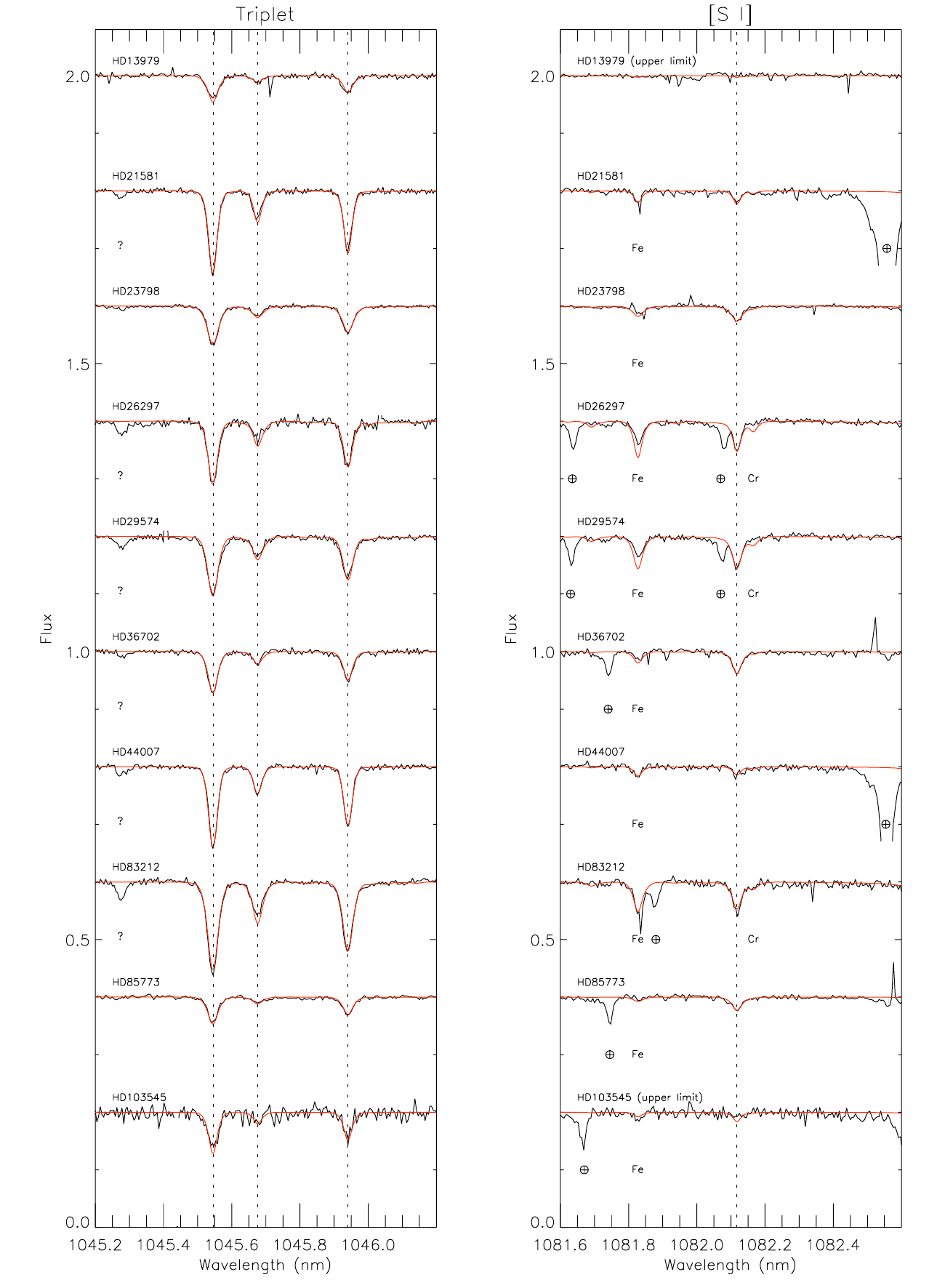}
\label{fig:forbidden}
\caption{The observed spectra in black and synthetic spectra based on the equivalent width analysis in red. The normalized spectra are shifted by multiples of 0.2 along the flux-axis in the figure. The downward spikes sometimes occurring, e.g. close to the Fe line in HD83212, are due to bad pixels. }
\label{fig:allspectra}
\end{figure*}

We have constructed stellar atmosphere models in spherical geometry for the observed stars using the MARCS code \citep{Gustafsson2008} and the stellar parameters quoted in Table \ref{tab:stellarparams}. In all models we use an $\left[\mathrm{\alpha}/\mathrm{Fe}\right]$ value of +0.4. The MARCS code computes hydrostatic model photospheres under the assumption of LTE (Local Thermodynamic Equilibrium), chemical equilibrium, homogeneity and the conservation of the total flux.
We used these models and the Uppsala Eqwi code (ver. 7.06) together with our measured equivalent widths (see Sect. \ref{eqwi}) to find two sulphur abundances for every star; one mean value based on the \ion{S}{i} 1045 nm triplet and one based on the [\ion{S}{i}] 1082 nm line. The observed and synthetic spectra with sulphur abundances determined in the way described can be seen in Fig. \ref{fig:allspectra}. The synthetic spectra are calculated with the Bsyn code (ver. 7.09), which is based on the same routines as the MARCS code. The continuous opacities are taken from the model atmospheres. Subsequently the spectra were broadened by convolving with a rad-tan function with a FWHM given in Table \ref{tab:stellarparams}, in order to fit the data taking into account the macroturbulence of the stellar atmosphere and instrumental broadening.

\subsection{Atomic data}
The atomic data for the line list used in the model calculations are taken from the VALD database \citep{Kupka1999}, except the data for the \ion{S}{i} triplet and [\ion{S}{i}] line which are taken from other sources, see Table \ref{tab:atomicdata}.
The [\ion{S}{i}] line is a M1 intercombination transition between the triplet ground state and the first excited singlet state. The 1045 nm triplet is three relatively highly exited transitions in the triplet system. 

\begin{table}[htp]
\caption{Atomic data for the relevant sulphur transitions.}
\begin{tabular}{c c c c c c}
\hline
\hline
Element  	&  Wavelength 	& $\chi_{\mathrm{exc}}$ 	& $\log (gf)$  	&  Transition & Refs.\\
 		&  (nm) (air) 	& (eV )				&			&\\
\hline
\ion{S}{i} 					&  1045.5449 	& 6.860 	& 0.250 	& $^3\mathrm{S}^{\mathrm{o}}_{1}\mathrm{-} ^3\mathrm{P}_{2}$ 	& 1\\
\ion{S}{i} 					&  1045.6757 	& 6.860 	& -0.447  	& $^3\mathrm{S}^{\mathrm{o}}_{1}\mathrm{-} ^3\mathrm{P}_{0}$ 	& 1\\
\ion{S}{i} 					&  1045.9406 	& 6.860 	& 0.0285  	& $^3\mathrm{S}^{\mathrm{o}}_{1}\mathrm{-} ^3\mathrm{P}_{1}$ 	& 1\\
$\left[\mathrm{\ion{S}{i}}\right]$ & 1082.1176 	& 0.000 	& -8.704 	& $^3\mathrm{P}_{2}\mathrm{-} ^1\mathrm{D}_{2}$ 				& 2\\
\hline
\end{tabular}
\label{tab:atomicdata}\\
\tablebib{(1) \citet{Zerne1997}; (2) \citet{Froese2011}}
\end{table}

We note that the $\log (gf)$-value of the [\ion{S}{i}] line is debated \citep{Asplund2009,Caffau2010a}, and the derived $\left[\mathrm{S}/\mathrm{Fe}\right]$ values would typically be 0.08 dex lower using the older $\log (gf)$-value used in \cite{Caffau2010a} (they use $\log (gf)=-8.617$).

\subsection{Stellar parameters}\label{stellarparams}
The observed stars all have Hipparcos parallaxes, but unfortunately the uncertainties for most of our observed stars are very large, see Table \ref{tab:starinfo}. This, in combination with the narrow wavelength coverage of CRIRES makes it impossible to determine the stellar parameters based on our observations alone, so we are forced to use stellar parameters from the literature. The stellar parameters were taken from \citet{Fulbright2003}, \citet{Gratton2000}, and \citet{Burris2000} (in turn mostly taken from \citet{Pilachowski1996}) and in case a star is listed in two or more of those references the newest is chosen.

\cite{Fulbright2003} obtained $T_{\mathrm{eff}}$ from colours, $\left[\mathrm{Fe}/\mathrm{H}\right]$ from \ion{Fe}{ii} lines, $\xi_{\mathrm{micro}}$ from forcing all $\ion{Fe}{i}$ line strengths to give the same iron abundance, and $\log g$ from the standard formula:\\
\begin{equation}
\log\frac{g}{g_{\odot}}=\log\frac{M}{M_{\odot}}+0.4\left(M_V+BC-4.72\right) +4\log\frac{T_{\mathrm{eff}}}{T_{\odot}}
\label{eq:logg}
\end{equation} \\
determining absolute magnitudes from literature values \citep{Hanson1998,Anthony-Twarog1994} in combination with fitting to a 12 Gyr isochrone and, where applicable, by using Hipparcos parallaxes. \cite{Gratton2000} determined $T_{\mathrm{eff}}$ from colours, $\log g$ from the standard formula (Equation \ref{eq:logg}) with absolute magnitudes from \cite{Anthony-Twarog1994}, $\xi_{\mathrm{micro}}$ from elimination of any trend for abundances as derived from $\ion{Fe}{i}$ lines with different strengths, and $\left[\mathrm{Fe}/\mathrm{H}\right]$ from \ion{Fe}{i} and \ion{Fe}{ii} lines. \cite{Burris2000} determined  $\left[\mathrm{Fe}/\mathrm{H}\right]$ from \ion{Fe}{ii} lines, but took the rest of the parameters from \cite{Pilachowski1996} who obtained $T_{\mathrm{eff}}$ from colours, $\xi_{\mathrm{micro}}$ from an iterative process fitting iron and calcium lines, and $\log g$ from three independent measurements: the standard formula (Equation \ref{eq:logg}) with absolute magnitudes from the Str\"omgren $c_1$-index, by assuming that the star's position in the M92 color-magnitude diagram, and through a derived average relation between surface gravity and effective temperature for metal-poor giants. The adopted stellar parameters are listed in Table \ref{tab:stellarparams} where we also give the uncertainties from the references. We note the slight inconsistency in adopting $\left[\mathrm{Fe}/\mathrm{H}\right]$-values as determined by plane-paralell model photospheres while we use spherical models.

In our observed wavelength region there are four \ion{Fe}{i}-lines which we used to check the $\left[\mathrm{Fe}/\mathrm{H}\right]$-values, given the other three parameters from the references are correct. Our $\left[\mathrm{Fe}/\mathrm{H}\right]$-values indeed fall within the uncertainties stated in the references. As can be seen in Fig. \ref{fig:allspectra} the Fe and Cr lines in the vicinity of the $\left[\ion{S}{i}\right]$ line, do not fit well with the given parameters for the two stars HD26297 and HD29574. Changing the $T_{\mathrm{eff}}$ and  $\left[\mathrm{Fe}/\mathrm{H}\right]$ within the uncertainty stated in \cite{Fulbright2003} will however make them fit. Despite this mismatch we have chosen to use the parameters determined by \cite{Fulbright2003} since their analysis is based on more and better known iron lines.

\begin{table*}[htp]
\caption{Adopted stellar parameters for the observed stars.}
\begin{tabular}{l c c c c c c}
\hline
\hline
Star & $T_{\mathrm{eff}}$ & $\log g$ & $\left[\mathrm{Fe}/\mathrm{H}\right]$ &  $\xi_{\mathrm{micro}}$ & Refs.& $\xi_{\mathrm{macro}}$\tablefootmark{b}\\
 & (K) & (cgs) & & (km s$^{-1}$) & & (km s$^{-1}$)\\
\hline
HD13979    &  $5075 \pm 100$			& $1.90 \pm 0.5$ 		& $-2.26 \pm 0.09$ 		& $1.30 \pm 0.5$ 		& 1	& 7.0\\
HD21581    &  $4900 \pm 103$ 		& $2.24 \pm 0.2$ 		& $-1.64 \pm 0.09$ 		& 1.45\tablefootmark{a}   	& 2	& 4.8\\ 
HD23798    &  $4375 \pm 82$   		& $1.12 \pm 0.2$ 		& $-2.03 \pm 0.13$ 		& 2.20\tablefootmark{a}  	& 2	& 7.3\\
HD26297    &  $4350 \pm 138$ 		& $1.46 \pm 0.2$ 		& $-1.51 \pm 0.11$ 		& 1.65\tablefootmark{a} 	& 2	& 5.4\\
HD29574    &  $4200 \pm 119$ 		& $0.78 \pm 0.2$ 		& $-1.70 \pm 0.12$ 		& 2.00\tablefootmark{a} 	& 2	& 6.0\\
HD36702    &  $4366 \pm 100$ 		& $0.95 \pm 0.2$ 		& $-2.06 \pm 0.11$		& $1.65 \pm 0.15$		& 3	& 6.0\\
HD44007    &  $4975 \pm 103$ 		& $2.24 \pm 0.2$ 		& $ -1.65 \pm 0.23$ 		& 2.20\tablefootmark{a}  	& 2	& 4.5\\
HD83212    &  $4533 \pm 100$ 		& $1.45 \pm 0.2$ 		& $ -1.40 \pm 0.13$ 		& $1.80 \pm 0.15$  		& 3	& 6.0\\
HD85773    &  $4450 \pm 100$ 		& $1.10 \pm 0.5$ 		& $ -2.36 \pm 0.09$ 		& $2.10 \pm 0.5$  		& 1	& 6.5\\
HD103545  & $4725 \pm 100$ 		& $1.70 \pm 0.5$ 		& $ -2.14 \pm 0.09$  		& $1.30 \pm 0.5$  		& 1	& 5.5\\
\hline
\end{tabular}
\label{tab:stellarparams}
\tablebib{(1)~\citet{Burris2000}; (2) \citet{Fulbright2003}; (3) \citet{Gratton2000}}
\tablefootmark{a}{The reference does not clearly state the uncertainty, so we assume a generic uncertainty of $\pm 0.5$ km s$^{-1}$. The size of this uncertainty does not affect the derived sulphur abundance much, see Table \ref{tab:stellarerror}, since the sulphur lines used in this work are weak and do not depend much on microturbulence.}\\
\tablefootmark{b}{The macroturbulence is not taken from the references, but determined by us to fit the synthetic spectra to the observed spectra in Fig. \ref{fig:allspectra}.}
\end{table*}

Typical uncertainties for the derived sulphur abundances due to typical uncertainties in the stellar parameters used are listed in Table \ref{tab:stellarerror}. As can be seen the 1045 nm triplet is most sensitive to the temperature, whereas the $[\ion{S}{i}] $ line, coming from a transition within the ground configuration, is not affected as much by uncertainties in the effective temperature, but more to uncertainties in the surface gravity through the dependence of the H$^-$ bound-free opacity on electron pressure, and the iron abundance. All the measured sulphur lines are weak and therefore not much affected by the microturbulence.

\begin{table}[htp]
\caption{A representative example (HD23798) of the effects on the sulphur abundances due to uncertainties in the stellar parameters.} 
\begin{tabular}{l c c}
\hline
\hline
Uncertainty & $\Delta \log \epsilon(\mathrm{S})_{\mathrm{triplet}}$ & $\Delta \log \epsilon(\mathrm{S})_{[\ion{S}{i}]}$\\
\hline
$\delta T_{\mathrm{eff}}=+100$ K 				& $-0.11$ 		& $+0.04$ \\
$\delta \log g= +0.2$ 						& $+0.08$  	& $+0.09$ \\  
$\delta\left[\mathrm{Fe}/\mathrm{H}\right]= +0.15$ 	& $+0.03$		& $+0.05$ \\ 
$\delta \xi_{\mathrm{micro}}= +0.5$ 				& $-0.03$  	&$\pm0.00$ \\ 
\hline
\end{tabular}
\label{tab:stellarerror}
\end{table}

\subsection{Non-LTE effects}
According to \cite{Takeda2005} the non-LTE corrections for the 1045 nm triplet are large and negative, i.e. the LTE abundances should be diminished in order to correct for the effects. They have calculated non-LTE corrections for a grid of model atmospheres with $T_{\mathrm{eff}}=4500, 5000, 5500, 6000, 6500, 7000$ K, $\log g=1.0, 2.0, 3.0, 4.0, 5.0$, and $\left[\mathrm{Fe}/\mathrm{H}\right]=-4.0, -3.0, -2.0, -1.0, -0.5, 0.0, +0.5$. Their calculations use hydrogen collision cross sections based on the classical formula by \citet{Drawin1968,Drawin1969} in the version of \citet{Steenbock1984} i.e. $S_H=1$ is assumed. The work of \cite{Takeda2005} is, to our knowledge, the only available non-LTE corrections for the 1045 nm triplet. \cite{Korotin2009} has, however, also performed non-LTE calculations for sulphur, but for the multiplet around 675 nm, the 869 nm doublet, and the 923 nm triplet. The overlapping parts of the two works agree well for lower-temperature stars, but differ for temperatures above 6000 K.

The non-LTE correction used \citep{Takeda2005} for our observed stars is shown in Table \ref{tab:abund}. Note that we have not attempted to interpolate in the grid, but chosen the corrections from the models best applicable to the stars in this study. We note the slight inconsistency of using spherical model photospheres for our giants while applying non-LTE corrections calculated in plane parallel geometry. Further investigations on the effects of non-LTE corrections due to spherical geometry would be useful.

Regarding the 1082 nm $[\ion{S}{i}] $ line, being an inter-combination resonance line, LTE is expected to be a good approximation, because the populations of the energy levels are determined by collisions rather than radiative transitions. 

\subsection{3D modeling}
We have used the set of 3D hydrodynamic model atmospheres of red giants by \citet{Collet2007,Collet2009} to perform spectral line formation calculations for the [$\ion{S}{i}$] line and the triplet. These calculations were compared to analogous 1D calculations based on MARCS models to estimate the 3D$-$1D LTE corrections to the sulphur abundance for giants with different stellar parameters in order to check the effects of our neglect of convective inhomogeneities on our abundance values. The results are shown in Table \ref{tab:3D}.

\begin{table}[htp]
\caption{3D - 1D corrections for $\log \epsilon(\mathrm{S})$.}
\begin{tabular}{ c c c c c }
\hline
\hline
$T_{\mathrm{eff}}$ & $\log g$ & $\left[\mathrm{Fe}/\mathrm{H}\right]$ & 3D$-$1D$_{[\ion{S}{I}]}$ & 3D$-$1D$_{\mathrm{triplet}}$ \\
 (K) & (cgs)\\
\hline
4717 & 2.2 & -1.0 & -0.04 & +0.17\\
5131 & 2.2 & -1.0 & -0.09 & +0.08\\
4732 & 2.2 & -2.0 & -0.08 & +0.19\\
5035 & 2.2 & -2.0 & -0.15 & +0.20\\
4858 & 2.2 & -3.0 & -0.27 & +0.20\\
5130 & 2.2 & -3.0 & -0.33 & +0.25\\
4550 & 1.6 & -3.0 & -0.09 & +0.17\\
\hline
\end{tabular}
\label{tab:3D}
\end{table}

The parameters of the 3D models are not exactly the same as our sample stars, but these calculations provide a rough estimate of the 3D-1D differences and some qualitative conclusions can be drawn: the corrections for the $[\ion{S}{i}] $ line seem to grow larger for higher temperatures, larger surface gravity and lower $\left[\mathrm{Fe}/\mathrm{H}\right]$. For our sample stars the 3D corrections for the sulphur abundances as derived from the $[\ion{S}{i}] $ line all seem to be around or below -0.1 dex.  The corrections for the triplet, on the other hand, are rather constant for the different 3D models and all seem to be around +0.2 dex. 

Our 3D analysis for these sulphur lines is the first for giants, but \cite{Caffau2007a} explored the 3D correction for the $[\ion{S}{i}]$ line in the Sun and \cite{Caffau2007b,Caffau2010} explored the 3D corrections for the 1045 nm triplet in the Sun, Procyon and four other dwarfs. All three works used $\mathrm{CO}^5\mathrm{BOLD}$ 3D hydrodynamical atmospheres for calculating the 3D effects.  \cite{Caffau2007a} found that the negative 3D-corrections for the $[\ion{S}{i}]$ line gets more significant (up to roughly $-0.2$ dex) for lower $\left[\mathrm{Fe}/\mathrm{H}\right]$ and \cite{Caffau2010} found positive 3D-corrections for the 1045 nm triplet more or less canceling the non-LTE corrections applied ($\sim+0.1$ dex). It is of course difficult to compare our models to the models used in \cite{Caffau2007a} and \cite{Caffau2007b, Caffau2010} because of the large difference in stellar parameters, but it can be noted, however, that all works predict a negative correction for the $[\ion{S}{i}] $ line and a positive for the triplet.

We have also estimated corrections to the Fe abundances taken from the literature (see Sect. \ref{stellarparams}), and found values from -0.1 dex to -0.2 dex for $\ion{Fe}{i}$ and from +0.05 dex to +0.1 dex for $\ion{Fe}{ii}$. 

\section{Results}
\subsection{Sulphur}
\begin{figure*}[htp]
\centering
\includegraphics[width=18cm]{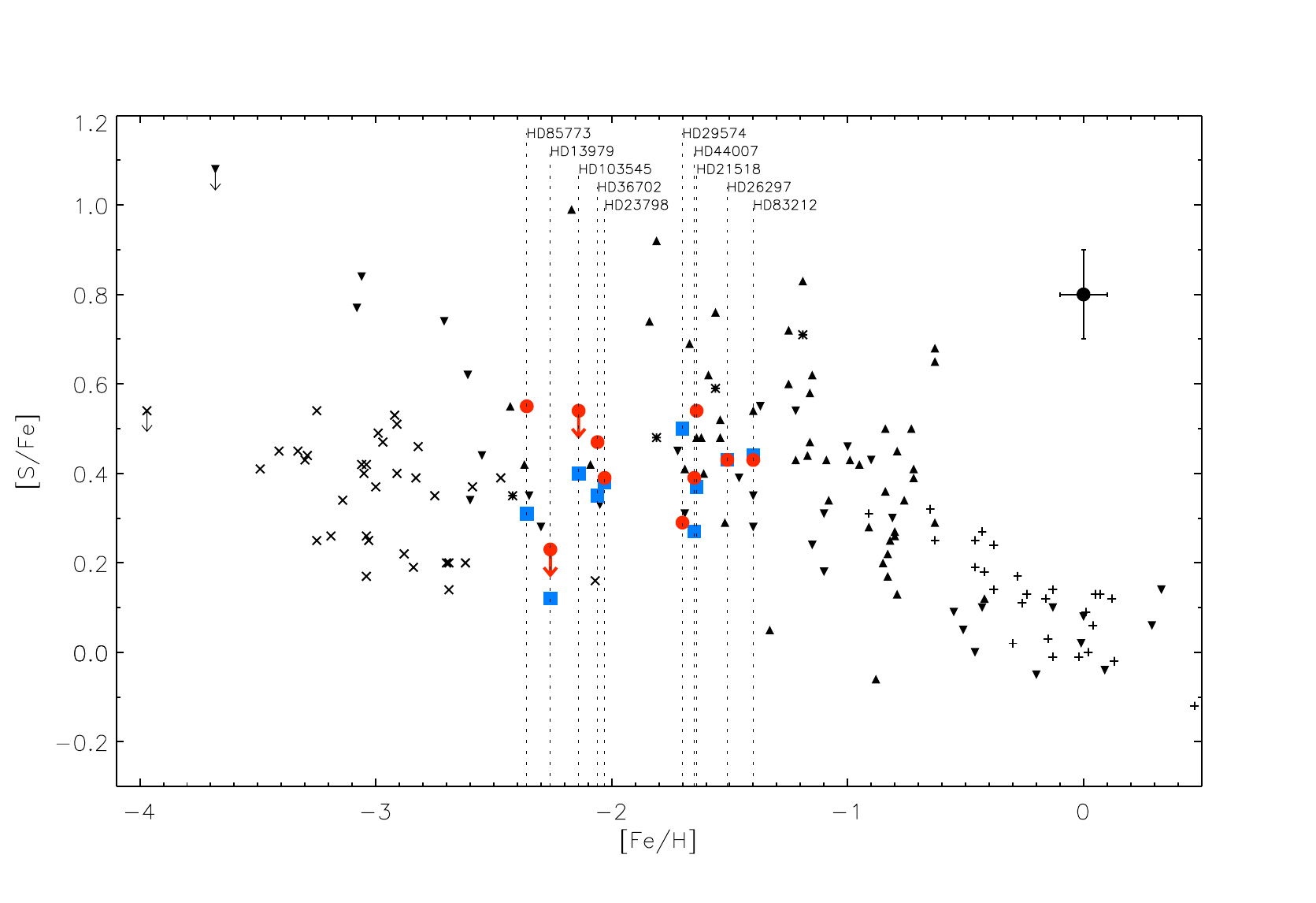}
\caption{$\left[\mathrm{S}/\mathrm{Fe}\right]$ versus $\left[\mathrm{Fe}/\mathrm{H}\right]$ for the values deduced from the $[\ion{S}{i}]$ line shown with red circles and the non-LTE corrected values deduced from the 1045 nm triplet shown with blue squares. Typical uncertainties based on quadratic addition of the values in Table \ref{tab:stellarerror} are shown in upper right hand corner. The plus signs are measurements from \citet{Chen2002}, upwards pointing triangles from \citet{Caffau2005}, asterixes from \citet{Caffau2010}, downwards pointing triangles from \citet{Takeda2010}, and exes from \citet{Spite2011}. They are all shifted to $\log \epsilon(S)_{\odot}=7.12$ \citep{Asplund2009}.}
\label{fig:evo}
\end{figure*}

The measured sulphur abundances and the applied non-LTE corrections from \cite{Takeda2005} can be seen in Table \ref{tab:abund}. In this table we also list the difference between the LTE and non-LTE sulphur abundances derived from the 1045 nm triplet and those derived from the 1082 nm $[\ion{S}{i}] $ line. The plot of   $\left[\mathrm{S}/\mathrm{Fe}\right]$ vs. $\left[\mathrm{Fe}/\mathrm{H}\right]$ can be seen in Fig. \ref{fig:evo} and \ref{fig:Si_evo}. We have used a solar sulphur abundance of $\log \epsilon(S)_{\odot}=7.12$ \citep{Asplund2009}  and all data in Fig. \ref{fig:evo} are put on this scale. A different choice of solar sulphur abundance would simply shift all data systematically by e.g., 0.04 dex using the value of \citet{Caffau2010a}.

In the Table \ref{tab:abund} and Figs. \ref{fig:evo} - \ref{fig:Si_evo} we have ignored 3D-corrections since we only have a coarse grid of models available for stars with parameters like our stars, but if they were to be applied they would work in the direction of lowering the sulphur abundances as determined from the [\ion{S}{i}]  line and adding to the abundances from the 1045 triplet.

\begin{table*}[htp]
\caption{Derived sulphur abundances.}
\begin{tabular}{l | c c c | r r | r r }
\hline
\hline
	Star & \multicolumn{3}{c}{Triplet} & \multicolumn{2}{c}{[\ion{S}{i}]} & \multicolumn{2}{c}{$\Delta \log \epsilon(\mathrm{S})_{\mathrm{triplet}-\left[\ion{S}{i}\right]}$}  \\
 & $\log \epsilon(\mathrm{S})$\tablefootmark{a} & NLTE\tablefootmark{b} &  $\left[\mathrm{S}/\mathrm{Fe}\right]_{\mathrm{NLTE}}$\tablefootmark{c} & $\log \epsilon(\mathrm{S})$ & $\left[\mathrm{S}/\mathrm{Fe}\right]$\tablefootmark{c} & LTE & NLTE\\

\hline
HD13979		&  5.13 $\pm 0.03$	& -0.15	& 0.12	& $\leq$ 5.09	& $\leq$ 0.23	& $\geq -0.04$	& $\geq-0.19$	\\
HD21581   	&  6.03 $\pm 0.05$	& -0.18 	& 0.37 	& 6.02 		& 0.54		& 0.01 		& $-0.17$	\\
HD23798   	&  5.65 $\pm 0.04$ 	& -0.18 	& 0.38 	& 5.48 		& 0.39		& 0.17		& $-0.01$	\\
HD26297   	&  6.20 $\pm 0.08$ 	& -0.16 	& 0.43 	& 6.04 		& 0.43		& 0.16		& $0.00$	\\
HD29574   	&  6.08 $\pm 0.09$ 	& -0.16 	& 0.50 	& 5.71 		& 0.29		& 0.37		& $0.21$	\\
HD36702   	&  5.57 $\pm 0.09$ 	& -0.16 	& 0.35 	& 5.53 		& 0.47		& 0.04		& $-0.12$	\\
HD44007   	&  5.89 $\pm 0.02$ 	& -0.15 	& 0.27 	& 5.86 		& 0.39		& 0.03		& $-0.12$	\\
HD83212   	&  6.44 $\pm 0.07$ 	& -0.28 	& 0.44 	& 6.15 		& 0.43		& 0.29		& $0.01$	\\
HD85773   	&  5.25 $\pm 0.05$ 	& -0.18 	& 0.31 	& 5.31 		& 0.55		& $-0.06$		& $-0.24$	\\
HD103545 	&  5.46 $\pm 0.06$ 	& -0.08 	& 0.40 	& $\leq$ 5.52 	& $\leq$ 0.54	& $\geq -0.06$	& $\geq-0.14$\\
\hline
\end{tabular}
\label{tab:abund}\\
\tablefootmark{a}{The uncertainty shows the line-to-line scatter between the three lines in the triplet.\\} 
\tablefootmark{b}{non-LTE corrections taken from \cite{Takeda2005}.\\} 
\tablefootmark{c}{Using a solar sulphur abundance of $\log \epsilon(S)_{\odot}=7.12$ \citep{Asplund2009}.}
\end{table*}

\subsection{Other alpha elements}
In our narrow wavelength range there are few lines suitable for determining abundances for other $\alpha$ elements, but there are three \ion{Si}{i} lines that can be used. The atomic data are taken from the NIST database and can be seen in Table \ref{tab:si_data}.

\begin{table}[htp]
\caption{Atomic data for the silicon transitions used.}
\begin{tabular}{c c c c c c}
\hline
\hline
Element  	&  Wavelength 	& $\chi_{\mathrm{exc}}$ 	& $\log (gf)$ 	&  Transition & Refs.\\
 		&  (nm) (air) 	& (eV )				&			&\\
\hline
\ion{Si}{i} 					&  1037.1264 	& 4.930 	& -0.712 	& $^3\mathrm{P}^{\mathrm{o}} \mathrm{-} ^3\mathrm{S}$ 	& 1,2\\
\ion{Si}{i} 					&  1084.3858 	& 5.862 	& 0.220  	& $^1\mathrm{P} \mathrm{-} ^1\mathrm{D}^{\mathrm{o}}$ 	& 1,3\\
\ion{Si}{i} 					&  1088.2809 	& 5.984 	& -0.646	& $^3\mathrm{D} \mathrm{-} ^3\mathrm{F}^{\mathrm{o}}$ 	& 1\\
\hline
\end{tabular}
\label{tab:si_data}
\tablebib{(1)~\citet{Radziemski1965};\\ (2)~\citet{Kelleher2008}; (3)~\citet{Nahar1993}} 
\end{table}

\begin{figure*}[htp]
\centering
\includegraphics[width=19 cm]{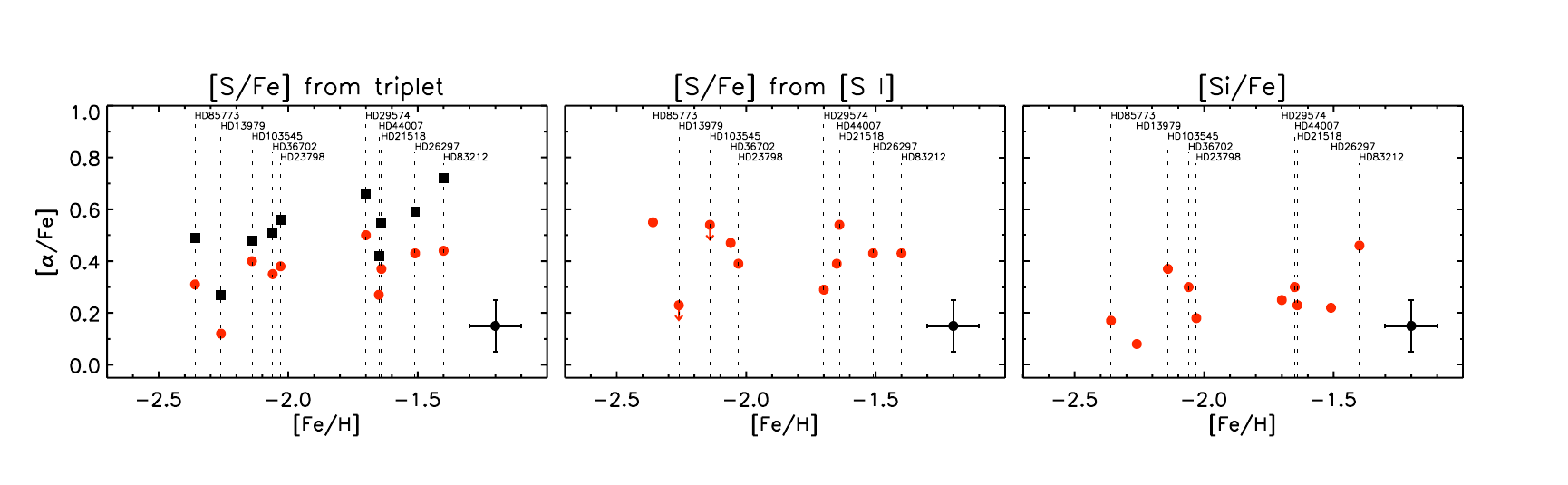}
\caption{$\left[\mathrm{S}/\mathrm{Fe}\right]$ and $\left[\mathrm{Si}/\mathrm{Fe}\right]$ versus $\left[\mathrm{Fe}/\mathrm{H}\right]$. The left panel shows LTE sulphur abundances deduced from the 1045 nm triplet in black squares and the non-LTE corrected values in red circles. The middle panel shows the sulphur abundances deduced from the 1082 nm $[\ion{S}{i}]$ line. The right panel shows the silicon abundance.}
\label{fig:Si_evo}
\end{figure*}

Equivalent widths for the \ion{Si}{i}-lines used are found in Table \ref{tab:si_eqw}. For stars with no values given for the 1037.1 nm and 1084.4 nm lines, no observations were made and for the two stars with no values given for the 1088.3 nm line, the line was covered by a telluric line. The derived silicon abundances are shown in Table \ref{tab:si_abund}, and the plot of  $\left[\mathrm{Si}/\mathrm{Fe}\right]$ vs. $\left[\mathrm{Fe}/\mathrm{H}\right]$ in Fig. \ref{fig:Si_evo}. We are not aware of any non-LTE calculations for these \ion{Si}{i}-lines.

\begin{table}[htp]
\caption{Measured equivalent widths for the observed silicon lines.}
\begin{tabular}{l c c c c c}
\hline
\hline
Star 	& 1037.1 nm 	& 1084.4 nm 	& 1088.3 nm \\
	& (m\AA) 		& (m\AA) 		& (m\AA)\\
\hline
HD13979   	& ...			& 22 $\pm$ 1       	& $\sim$ 4    	\\
HD21581   	& 83 $\pm$ 1    	& ...			  	& ... 			\\    
HD23798   	& ...  			& 58 $\pm$ 1            	& 12 $\pm$ 1 	\\
HD26297   	& ...     		& 86 $\pm$ 1     	& 24 $\pm$ 1	\\
HD29574   	& ...     		& 88 $\pm$ 1      	& 21 $\pm$ 1  	\\
HD36702   	& ...  			& 63 $\pm$ 1	    	& 14 $\pm$ 2 	\\
HD44007   	& 84 $\pm$ 1	& 89 $\pm$ 1  		& 27 $\pm$ 2 	\\
HD83212   	& ...     		& 123 $\pm$ 1 		& 44 $\pm$ 2 	\\
HD85773   	& ...  			& 39 $\pm$ 1           	& $\leq$ 5 	\\
HD103545 	& 57 $\pm$ 2    	& 63 $\pm$ 1   		& ...    		\\
\hline
\end{tabular}
\label{tab:si_eqw}
\end{table}

\begin{table}[htp]
\caption{Derived silicon abundances.}
\begin{tabular}{l c c}
\hline
\hline
Star & $\log \epsilon(\mathrm{Si})$ & $\left[\mathrm{Si}/\mathrm{Fe}\right]$\tablefootmark{a}\\
\hline
HD13979		&  5.29 	& 0.04 \\	
HD21581   	&  6.06 	& 0.19 \\
HD23798   	&  5.62 	& 0.14 \\
HD26297   	&  6.18 	& 0.18 \\
HD29574   	&  6.02 	& 0.21 \\
HD36702   	&  5.71 	& 0.26 \\
HD44007   	&  6.12 	& 0.26 \\
HD83212   	&  6.53 	& 0.42 \\
HD85773   	&  5.28 	& 0.13 \\
HD103545 	&  5.70 	& 0.33 \\
\hline
\end{tabular}
\label{tab:si_abund}\\
\tablefootmark{a}{Using a solar silicon abundance of $\log \epsilon(Si)_{\odot}=7.51$  \citep{Asplund2009}.}
\end{table}

\section{Discussion}
Until a decade ago, only few studies of sulphur abundances in Galactic halo stars (or, more correctly, for stars more metal-poor than [Fe/H]$<-1$),  had been made, mainly due to the lack of suitable diagnostic lines. The weak  doublet around 869 nm, a transition from a relatively high level at 7.9 eV, was used but these lines are increasingly  difficult to measure as the metallicity decreases. However their formation is close to LTE conditions, and are therefore a recommended diagnostic when detactable \citep{Takeda2005}. In 2004 Nissen et al. and Ryde \& Lambert  used the triplet lines around 923 nm ($\chi_{exc}=6.5$ eV) which are strong enough for determining  sulphur abundances at high precision in halo stars, also for the metal-poor ones. The 923 nm triplet is heavily affected by telluric lines, but this can nicely be handled with an observation of a telluric standard star \citep{Ryde2004, Nissen2004}. A severe problem with these stronger lines has been that not all optical CCDs reach up to the wavelengths required. Also, the lines are presumably not formed in LTE \citep{Takeda2005}.
Recently, the diagnostic power of new sulphur lines in the near IR, but beyond the reach of normal CCDs, have been explored by means of spectrometers like CRIRES at VLT and near-IR detector arrays \citep[i.e.,][]{Ryde2006,Nissen2007}. These lines are the 1082 nm [$\ion{S}{i}$] line and 1045 nm triplet explored and used in this work. The 1045 nm triplet has non-LTE corrections calculated by \citet{Takeda2005} and the [$\ion{S}{i}$] line, being a forbidden line, is formed under close to LTE conditions, which is confirmed by non-LTE calculations by Korotin (private communication). This makes analyses based on other sulphur lines than the 869 nm doublet or the 1082 nm [$\ion{S}{i}$] line rely on more or less uncertain\footnote{It should be noted that non-LTE calculations are subject to a number of more or less important uncertainties and assumptions. The non-LTE corrections are therefore also uncertain.}, but necessary,  non-LTE calculations. 

In this paper we have analyzed ten halo giants observed in the near-IR at high spectral resolution in order to determine the sulphur abundances from the near-IR triplet at 1045 and the forbidden line at 1082 nm. We are able, for the first time, to analyse the 1082 nm [$\ion{S}{i}$] line for halo stars and show that it is indeed a useful diagnostic in giants down to at least $\left[\mathrm{Fe}/\mathrm{H}\right] \sim -2.3$. This result is in agreement with our model predictions (see Fig. \ref{fig:SI_eqw}) showing that it should be usable even for lower $\left[\mathrm{Fe}/\mathrm{H}\right]$ for cool stars with low surface gravity. Our model predictions also show that in dwarfs and subgiants this line is only detectable down to $\left[\mathrm{Fe}/\mathrm{H}\right] \sim -1$ with the present means. 

Our results from the $[\ion{S}{i}] $ line favor a flat trend for $\left[\mathrm{S}/\mathrm{Fe}\right]$ as a function of $\left[\mathrm{Fe}/\mathrm{H}\right]$ for halo stars (see Fig. \ref{fig:evo} and \ref{fig:Si_evo}). Fitting a line to the measurements\footnote{We used the least square linear fit implemented in the routine \texttt{fitexy} presented in NASA's IDL library at \href{http://idlastro.gsfc.nasa.gov/}{http://idlastro.gsfc.nasa.gov/}. The routine takes the uncertainties in both $\left[\mathrm{S}/\mathrm{Fe}\right]$ and $\left[\mathrm{Fe}/\mathrm{H}\right]$ into account.}  
gives $\left[\mathrm{S}/\mathrm{Fe}\right]=-0.021 \cdot \left[\mathrm{Fe}/\mathrm{H}\right] +0.39$ and the simple mean gives $\left[\mathrm{S}/\mathrm{Fe}\right]=0.43$ with a standard deviation of $\sigma = 0.11$. Since the $[\ion{S}{i}] $ line is not believed to be affected by non-LTE effects, and 3D effects are smaller, we consider these measurements more robust than the measurements using the 1045 nm triplet. By just comparing the two leftmost panels in Fig. \ref{fig:Si_evo} it seems that the non-LTE corrected sulphur abundances from the 1045 nm triplet are similar, but perhaps somewhat lower, than the abundances from the 1082 nm [\ion{S}{i}]  line.
We have conducted a Kolmogorov-Smirnov test\footnote{We used the routine  \texttt{kstwo} from \cite{Press1992}.} checking the probability for all our non-LTE corrected $\left[\mathrm{S}/\mathrm{Fe}\right]$ 1045 nm triplet measurements to come from the $\left[\mathrm{S}/\mathrm{Fe}\right]$ distribution as described by our $[\ion{S}{i}]$ line measurements. The Kolmogorov-Smirnov statistic for these two distributions is 0.40 and p-value for this is 0.31\footnote{The Kolmogorov-Smirnov statistic gives the maximum deviation between the cumulative distribution function of the triplet sulphur abundances from the values deduced from the $[\ion{S}{i}]$-line. The p-value takes values between 0 and 1 giving the significance level of the K-S statistic. A large p-value would, in our case, show that the cumulative distribution function of our non-LTE corrected $\left[\mathrm{S}/\mathrm{Fe}\right]$ 1045 nm triplet measurements is similar to the cumulative distribution function of our $\left[\mathrm{S}/\mathrm{Fe}\right]$ $[\ion{S}{i}] $ line measurements.}, meaning that formally the two distributions are similar to a low degree and therefore pointing toward a possible problem with the non-LTE corrections applied to the 1045 nm triplet. When instead comparing the sulphur abundance distribution from the $[\ion{S}{i}]$-line with the \emph{LTE} sulphur abundance distribution from the triplet the Kolmogorov-Smirnov statistic is as expected larger; 0.50 and the p-value 0.11. Thus the non-LTE corrections of \cite{Takeda2005} at least adjust the LTE distribution of sulphur abundances from the triplet to be \emph{more} like our resulting abundances from the $[\ion{S}{i}]$-line.

In our analysis we have so far ignored 3D-corrections due to the lack of models available for stars with parameters like our stars. It is of course very hard to inter- and extrapolate our coarse 3D-grid to give us information applicable to the stars observed, but it would seem that the positive 3D corrections are large enough (almost +0.2 dex) to cancel the applied non-LTE corrections making the sulphur abundance measured from the 1045 nm triplet higher than the abundances from the 1082 nm [$\ion{S}{i}$] line. This would make the differences between the two data sets seen in Fig. \ref{fig:Si_evo} and in the results of the Kolmogorov-Smirnov test larger again and this \emph{might} indicate that the non-LTE corrections of \cite{Takeda2005} are too small, or that our estimation of the 3D corrections is too large. From Fig. \ref{fig:evo} it indeed seems like the agreement between the two diagnostics would be even better if the [S/Fe] deduced from the [\ion{S}{i}] line were lower and the values deduced from the triplet were higher, which at least coincides with the \emph{direction} of the 3D corrections.  It is however expected that non-LTE corrections fully incorporating convective inhomogeneities would be different than the 1D non-LTE corrections used. This needs further study with the coupling between 3D and non-LTE done self-consistently. {We note that when removing the applied 3D-corrections from \citet{Caffau2010} three of their four measurements, using the 1045 nm triplet, (asterixes in Fig. \ref{fig:evo}) overlap very well with our 1045 nm triplet measurements. This is interesting since the measurements were made using the same instrument, the same diagnostic, and the same non-LTE corrections, but \citet{Caffau2010}Êused dwarfs instead of giants like we do. This might be another indication that the non-LTE corrections work well in 1D.

Using the [\ion{S}{i}]  line as a diagnostic for determining sulphur abundance in halo red giants is preferable due to its small non-LTE and 3D effects, but using \emph{only} this line is  problematic, not only as exemplified by our data where an instrumental ghost  unfortunately hit our record of this line, but also because it might be blended by \emph{unidentified} stellar lines, or telluric lines. Fortunately we cannot find any blends and, concerning the ghost, we found a way to rescue the data and determine the sulphur abundance from this line. Because of this risk more lines should be used simultaneously, preferably both the 923 nm and 1045 nm non-LTE corrected triplets. It should be noted that the non-LTE formation of these lines are still not well understood \citep{Takeda2005}. Observing both the 923 nm triplet and the 1082 nm [\ion{S}{i}] line would provide the same test of the non-LTE corrections of the 923 nm triplet as presented here for the 1045 nm triplet.

We find that the uncertainties in $\left[\mathrm{S}/\mathrm{Fe}\right]$ in this study will mainly be due to uncertainties in the stellar parameters (in our case $\sim 0.1$ dex), the continuum drawing, (in our case $< 0.1$ dex) and due to non-LTE corrections for the triplet and effects from convective inhomogeneities. Given these uncertainties no significant abundance trend in $\left[\mathrm{S}/\mathrm{Fe}\right]$ vs. $\left[\mathrm{Fe}/\mathrm{H}\right]$ but a scatter around  $\left[\mathrm{S}/\mathrm{Fe}\right] \sim 0.4$ can be inferred from our data. To determine whether this scatter is real we investigated the correlation between $\left[\mathrm{S}/\mathrm{Fe}\right]_{\mathrm{triplet}}$ and $\left[\mathrm{S}/\mathrm{Fe}\right]_{[\ion{S}{i}]}$. They would be correlated if the scatter was real, but such a trend would also result from errors in the $\left[\mathrm{Fe}/\mathrm{H}\right]$ used. We cannot find any correlation, but only scatter from what is expected from the uncertainties, except for HD13979. We note that if we use the newer parameters and iron abundance of \cite{Wylie2010} ($T_{\mathrm{eff}}=5050$ K, $\log g=2.25$, $\left[\mathrm{Fe}/\mathrm{H}\right]=-2.48$, $\xi_{\mathrm{micro}}=2.35$ km s$^{-1}$) we get the sulphur abundances of HD13979 to more resemble the trend of the others suggesting that our used $\left[\mathrm{Fe}/\mathrm{H}\right]$ from \citet{Burris2000} might not be correct.

\citet{Caffau2005} find $\sim 20$ \% of their analyzed stars to have `high' values of $\left[\mathrm{S}/\mathrm{Fe}\right]$ meaning that our randomly chosen sample of 10 stars would with 90 \% probability find at least one from this population, if real, but we do not. Concerning the zig-zag trend of \citet{Takeda2010} our sample of stars unfortunately do not go low enough in iron abundance to confirm or whether their rise is real or not. Our sample of stars do not have any stars in common with the previous studies to compare abundances from these with our [$\ion{S}{i}$] abundances.

The three \ion{Si}{i} lines available in our spectra give a flat silicon abundance trend with roughly equal behavior as the flat sulphur abundance trend in our observed stars, which is what is expected from their common nucleosynthetic origin. The sulphur abundance of HD13979 is low compared to the others, but also the silicon abundance is low, meaning that it is possible that this star \emph{is} low in abundance of $\alpha$ elements, but, as mentioned earlier, the stellar parameters used for this star might not be correct. If we use the newer parameters of \cite{Wylie2010} also the silicon abundance of HD13979 resemble the trend of the others.

As can be seen in Fig. \ref{fig:allspectra}, the Fe and Cr lines in the synthetic spectra of HD26297 and HD29574 do not fit the data. This is likely due to the uncertainties in the adopted stellar parameters and the $gf$-values of the lines. As noted earlier we still have chosen to use the parameters listed in \cite{Fulbright2003} since our analysis of $\left[\mathrm{Fe}/\mathrm{H}\right]$ would be based on Fe-lines with lower quality.

An increasing number of investigations, but not all, are gathering evidence toward a view where sulphur follows the abundance trends vs. metallicity in the halo expected for the $\alpha$ elements, i.e. at a constant value to low metallicities. This have implications for models of supernova and hypernova yields and can constrain such models. For example the hypernova yield models by \citet{Nakamura2001} used by \citet{Israelian2001} to explain their `high' values of $\left[\mathrm{S}/\mathrm{Fe}\right]$ are not confirmed in newer yield calculations by \citet{Kobayashi2006}, but they predict a flat $\left[\mathrm{S}/\mathrm{Fe}\right]$ vs. $\left[\mathrm{Fe}/\mathrm{H}\right]$ trend for halo stars in agreement with e.g. our results. However, the predictions of high $\left[\mathrm{S}/\mathrm{Fe}\right]$ by \citet{Nakamura2001} refer to hypernovae with explosions energy of $100 \times 10^{51}$ erg, whereas \citet{Kobayashi2006} only include hypernovae with energies up to $30 \times10^{51}$ erg, but it can be noted that \citet{Nakamura2001} also predict enhanced Si/O and Ca/O ratios for the most energetic hypernovae, which are not observed, so it seems that such very energetic hypernovae do not play any role for the chemical evolution above $\left[\mathrm{Fe}/\mathrm{H}\right] \sim -3$.

\section{Conclusion}
In the literature, there are several proposed trends for the evolution of sulphur in the halo phase of the Milky Way. This has implications for our understanding of the nucleosynthesis of sulphur and the sites where this occurs. The three currently discussed trends are the flat trend of e.g., \cite{Spite2011}, the scatter trend of \cite{Caffau2010}, and the recent zig-zag trend of \cite{Takeda2010}.

In trying to minimize the effects of non-LTE in the diagnostic we have used the 1082 nm [\ion{S}{i}] line for which LTE is a good approximation and found a flat trend for $\left[\mathrm{S}/\mathrm{Fe}\right]$  similar to that in \cite{Spite2011} and the trends expected for the $\alpha$-elements. We also determined the sulphur abundance from the 1045 nm triplet to investigate its usefulness in giants, especially concerning its sensitivity to non-LTE. Our empirical differences can be compared with  existing non-LTE corrections.  We find that with the non-LTE corrections from \cite{Takeda2005} for the 1045 nm triplet, the derived  sulphur abundance in most giants are closer to the abundance determined from the forbidden line, but the data hints towards these non-LTE corrections being too large when 1D models are used. When considering 3D we conclude that they on the other hand \emph{might} be too small. This, however, is based on mostly qualitative conclusions from a  coarse 3D model grid, and it might be that the discrepancy between the two data sets is due to exaggerated 3D effects and not due to faulty non-LTE corrections.

Not only since it is formed under LTE conditions, but also as the 3D corrections are smaller for the [\ion{S}{i}] line as compared to the 1045 nm triplet, the 1082 nm [\ion{S}{i}] line would be the preferred diagnostic of the two used in this paper.  Even though the two diagnostics do not produce the exact same trend, our sulphur abundance from the near-IR triplet, corrected for the calculated non-LTE effects, also corroborates the idea of sulphur behaving like a normal $\alpha$ element and we do not find any `high' values for $\left[\mathrm{S}/\mathrm{Fe}\right]$ like in the scatter trend of \citet{Caffau2010}. Neither do we find any `high' values for $\left[\mathrm{S}/\mathrm{Fe}\right]$ like in the zig-zag trend of \citet{Takeda2010}, but unfortunately our sample of giants does not go to $\left[\mathrm{Fe}/\mathrm{H}\right]$ low enough to sample the region where their values of $\left[\mathrm{S}/\mathrm{Fe}\right]$ rise.

We have demonstrated that both the 1045 nm triplet and the 1082 nm [\ion{S}{i}] line should be useful for investigations of the halo phase of the Milky Way. The [\ion{S}{i}] line has the advantage of not being subject to uncertain non-LTE corrections, but it also has the drawback of being a single line.  An increasing number of investigations point toward sulphur behaving as a traditional $\alpha$ element in the trends in $\left[ \alpha / \mathrm{Fe}\right]$ vs $\left[\mathrm{Fe}/\mathrm{H}\right]$ plots. However, there still exist considerable claims for higher values as the metallicity decreases. To convincingly determine the situation for the $\alpha$- element sulphur, further systematic and homogeneous investigations are needed using as many diagnostics and with as sophisticated analyses as possible (including non-LTE and 3D modeling).

\begin{acknowledgements}
This research has been partly supported by the Royal Physiographic Society in Lund, Stiftelsen Walter Gyllenbergs fond and M\"arta och Erik Holmbergs donation. Also support from the Swedish Research Council, VR, project number 621-2008-4245, is acknowledged. N.R. is a Royal Swedish Academy of Sciences Research Fellow supported by a grant from the Knut and Alice Wallenberg Foundation. K.E. gratefully acknowledge support from the Swedish Research Council.\\ This publication made use of the SIMBAD database, operated at CDS, Strasbourg, France and NASA's Astrophysics Data System. 
\end{acknowledgements}

\bibliography{sulphur}
\bibliographystyle{aa}

\end{document}